\documentclass[acmsmall]{acmart}
\setcopyright{none}

\usepackage{graphicx}
\usepackage{algorithm,algpseudocode}
\usepackage{longfbox}
\usepackage{amsmath}
\usepackage{amsmath,amsfonts,mathtools,mathrsfs}
\usepackage{epsfig,amstext,xspace}
\usepackage{amsthm}
\usepackage[sans]{dsfont}
\usepackage{stmaryrd}
\usepackage{pifont}
\usepackage{wasysym}
\usepackage{tabularx}
\usepackage{array}
\usepackage{url}
\usepackage{color}
\usepackage{subfigure}
\usepackage{caption}
\usepackage{enumitem}
\usepackage{multirow}
\usepackage{float}

\newcolumntype{P}[1]{>{\centering\arraybackslash}p{#1}}

\begin{document}

\title{Integrating IoT-Sensing and Crowdsensing with Privacy: Privacy-Preserving Hybrid Sensing for Smart Cities}

\author{Hanwei Zhu}
\affiliation{%
  \institution{Australian National University}
  \country{Australia}
}
\email{henry.zhu@anu.edu.au}

\author{Sid Chi-Kin Chau}
\affiliation{%
  \institution{Australian National University}
  \country{Australia}
}
\email{sid.chau@anu.edu.au}

\author{Gladhi Guarddin}
\affiliation{%
  \institution{Universitas Indonesia}
  \country{Indonesia}
}
\email{adin@cs.ui.ac.id}

\author{Weifa Liang}
\affiliation{%
  \institution{City University of Hong Kong}
  \country{Hong Kong}
}
\email{weifa.liang@cityu.edu.hk}
\renewcommand{\shortauthors}{H. Zhu et al.}

\begin{abstract}
Data sensing and gathering is an essential task for various information-driven services in smart cities. On the one hand, Internet of Things (IoT) sensors can be deployed at certain fixed locations to capture data reliably but suffer from limited sensing coverage. On the other hand, data can also be gathered dynamically through crowdsensing contributed by voluntary users but suffer from its unreliability and the lack of incentives for users' contributions. In this paper, we explore an integrated paradigm called ``{\em hybrid sensing}'' that harnesses both IoT-sensing and crowdsensing in a complementary manner. In hybrid sensing, users are incentivized to provide sensing data not covered by IoT sensors and provide crowdsourced feedback to assist in calibrating IoT-sensing. Their contributions will be rewarded with credits that can be redeemed to retrieve synthesized information from the hybrid system. In this paper, we develop a hybrid sensing system that supports explicit user privacy -- IoT sensors are obscured physically to prevent capturing private user data, and users interact with a crowdsensing server via a privacy-preserving protocol to preserve their anonymity. A key application of our system is smart parking, by which users can inquire and find the available parking spaces in outdoor parking lots. We implemented our hybrid sensing system for smart parking and conducted extensive empirical evaluations. Finally, our hybrid sensing system can be potentially applied to other information-driven services in smart cities.
\end{abstract}



\maketitle

\section{Introduction}

Smart cities improve the quality of life by offering information-driven services. For instance, intelligent transportation services can adapt to dynamic users' demands for service optimization. To enable responsive and efficient information-driven services, it relies critically on timely and comprehensive data sensing and gathering to synthesize the global knowledge of user usages, behaviors, and patterns for service planning and management. For example, one of the key information-driven services is to enable smart parking in open spaces. According to a recent survey \cite{suvey21parking}, searching for available parking spaces is one of the topmost crucial activities of city dwellers. Currently, a large number of parking spaces are located in open areas (e.g., streets and campuses), with limited knowledge of their instantaneous statuses. Without sufficient knowledge of available parking spaces, drivers often spend numerous hours cruising around and competing for limited parking spaces. A more effective solution is to gather the availability data of parking spaces in a temporal and spatial manner from diverse sources and then disseminate the knowledge of available parking spaces to users in a timely fashion.

There are two major approaches for effective data sensing and gathering in smart cities:
\begin{itemize}

\item {\bf IoT-sensing}: IoT sensors can be deployed at specific fixed locations to capture their surrounding data reliably. Among the variety of sensors available in the market (e.g., short-range motion detection and long-range visual detection), one of the most common media is vision-based capturing with visual images and videos for object classification and detection, which has been used in monitoring traffic congestions, parking availabilities, and pedestrian flows. Mounted cameras can overlook an ample open space and can be installed flexibly. More importantly, cheaper cameras with higher resolution are increasingly available, which improves the quality and reliability of the captured images. However, IoT sensors may suffer from limited sensing coverage due to either the positioning of the sensors or the amount of deployed sensors. Also, IoT sensors sometimes experience calibration and privacy issues.

\item {\bf Crowdsensing}: Data can be gathered through crowdsensing contributed by voluntary users. Users or user-owned sensors can provide their surrounding data in a dynamic fashion. For example, users can report their observations captured by their smartphones. Navigation apps, such as Waze and Google Maps, continuously collect users' mobility data to infer traffic conditions, predict traveling delays, and generate traffic alerts. Some mobile apps allow users to share personal observations, such as public transport waiting times, congestion statuses, parking availabilities, and petrol prices. However, crowdsensing is not always reliable because it is based on the voluntary contributions of users. There is no guarantee that the crowdsensed data is always accurate or up-to-date.  
\end{itemize}

Although IoT-sensing and crowdsensing are different, they are often complementary to each other in terms of coverage and sensing reliability. Hence, there is a potential to harness both approaches. In this article, we explore an integrated paradigm called {\em hybrid sensing}, by which both IoT-sensing and crowdsensing are utilized to enhance the accuracy, coverage, and reliability of data sensing and gathering. Specifically, the benefits of such integration include: 
\begin{enumerate}
    \item While IoT sensors are deployed at fixed locations, crowdsensing can offer additional dynamic sensing coverage.
    \item Since crowdsensing is often difficult to initiate due to limited user engagement, IoT-sensed data can be used to bootstrap the process and further attract user contributions.
    \item IoT-sensed data can be calibrated and enhanced based on crowdsourced user feedback as part of crowdsensing. 
\end{enumerate}

However, to realize the full potential of hybrid sensing, we need to address the following two major issues explicitly:
\begin{itemize}

\item {\bf Privacy}: While sensors are deployed to collect data, they are not supposed to identify individual people or vehicles. There are increasing concerns about privacy violations by these sensors. The images and video footages of pedestrians and vehicles may expose their identities and behaviors, which easily lead to unintended criminal consequences (e.g., stalking and theft). Even if the administrators of these vision-based sensor systems intentionally discard the private data, their systems may still be infiltrated by hackers for malicious purposes. Particularly, there have been several webcam-related hacking and data breach instances where cameras were fully controlled by hackers, and the video footages were leaked to the Internet. On the other hand, users are also wary of sharing private personal data. In particular, the breach of personal privacy (e.g., identity and location information) remains a top concern for data sharing with personal and mobile devices.

\item {\bf Incentives}:  Ideally, crowdsensing is expected to benefit users by offering global knowledge to them from large amounts of shared data and hence motivates more users to share their personal data to improve the quality of crowdsensing. However, in reality, users are not keen to contribute to crowdsensing because they sometimes can enjoy benefits without any contribution as free riders. Insufficient data sharing will lead to a degraded quality in crowdsensing, undermining the willingness of further voluntary contributions of users. Thus, there is a need to incentivize crowdsensing to provide high-quality user-contributed data. 

\end{itemize}

Remarkably, our hybrid sensing system addresses these issues in a complementary manner. First, our system supports explicit user privacy -- IoT sensors are obscured physically to prevent capturing private user data, and users interact with a crowdsensing server via a privacy-preserving protocol to preserve their anonymity. Second, users are incentivized to provide sensing data not covered by IoT sensors and provide additional crowdsourced feedback to assist in calibrating IoT-sensing. Their contributions will be rewarded with credits that can be redeemed to retrieve synthesized crowdsensing information from our system. 

Note that hybrid sensing is a broad concept that can be applied in various information-driven services in smart cities, such as reporting garbage disposals, traffic conditions, and wildlife sightings. In this paper, we focus on the application of smart parking to demonstrate the usefulness and practicality of our system. Particularly, our hybrid sensing smart parking system enables users to inquire and find geographically distributed available parking spaces. 

\begin{figure}[ht] 
  \centering
  \includegraphics[scale=0.4]{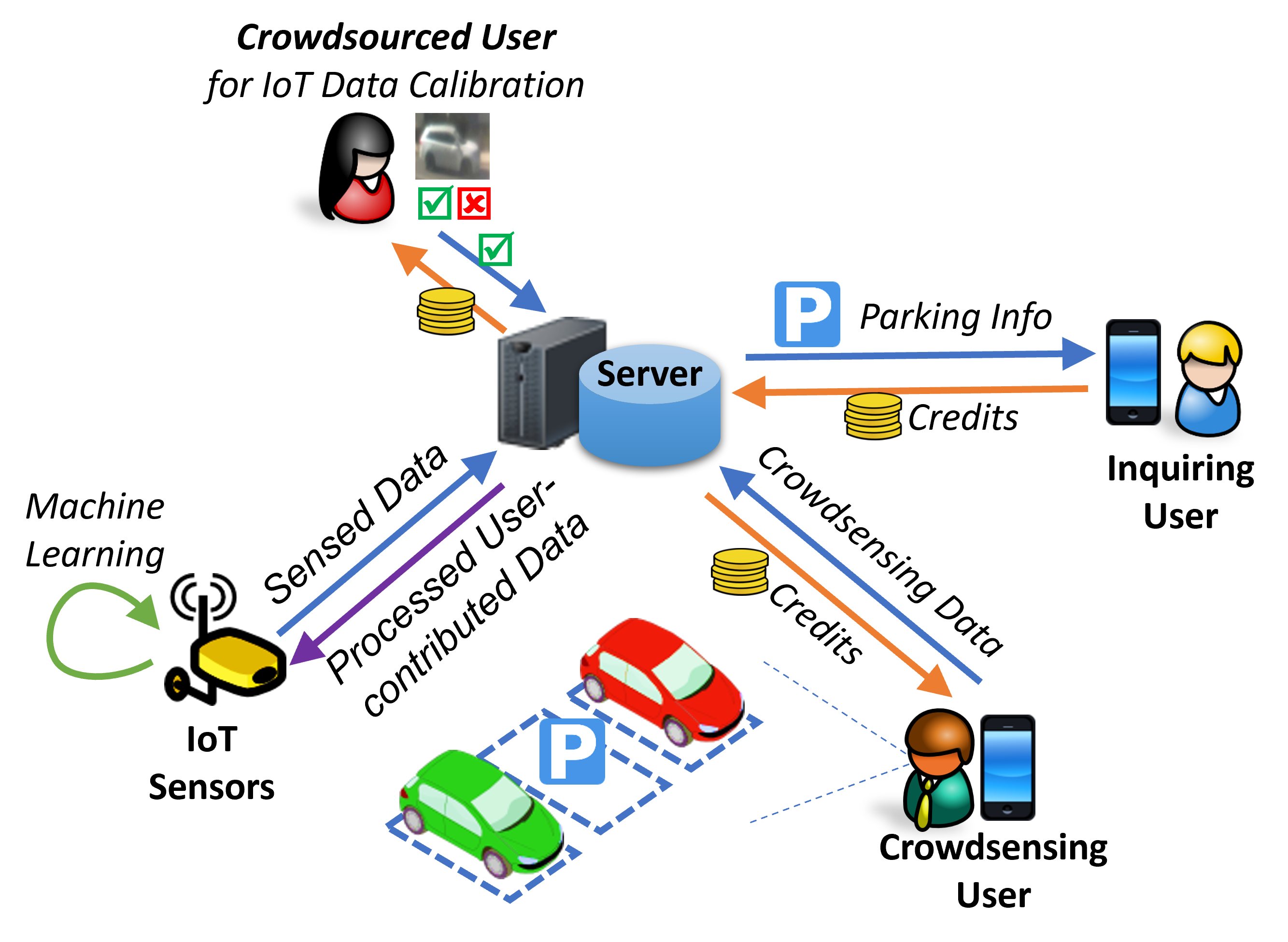}
  \caption{System design of a hybrid sensing system for smart parking.}
  \label{fig:scenario}
\end{figure}

As illustrated in Fig.~\ref{fig:scenario}, our smart parking system deploys Raspberry Pi-based fish-eye cameras for capturing the footages of parking spaces, which are obscured by blurry filters to enhance privacy. We developed an onboard machine learning system to recognize objects from captured blurry videos in real-time. On the one hand, complementary to the captured sensor data from the deployed cameras, users can also submit the observed available parking spaces through a mobile app via a privacy-preserving protocol using cryptographic commitments, zero-knowledge proofs, and anonymous credentials. 

On the other hand, to improve detection accuracy and to reduce the effort of training, in our machine learning system, we utilized crowdsourced classification data on segmented images from the captured footages to calibrate the machine learning model. The user-contributed data will then be validated by the server for usefulness. To incentivize user contributions, users will obtain credits for their valuable contributions while preserving their anonymity. The credits, as incentives, can be redeemed for parking inquiries. Since not every parking lot has been deployed with camera systems, crowdsensing will be particularly valuable to expand the coverage of IoT-sensing.

{\bf Organization:} The rest of this paper is organized as follows. We first review the related work in Section~\ref{sec:related}. Then, we present the privacy-preserving IoT-sensing part in Section~\ref{sec:IoT sensing} and the preliminaries of cryptographic components in Sections~\ref{sec:crypto}. Based on the cryptographic components, we develop our privacy-preserving crowdsensing protocol in Section~\ref{sec:crowdsensing}. Next, we describe our hybrid sensing system implementation in Section~\ref{sec:implementation}, and present an evaluation study of the implemented system in Section~\ref{sec:evaluation}. Finally, Section~\ref{sec:conclusion} concludes this paper. Some further analysis and technical proofs are included in the Appendix.
\section{Related Work} \label{sec:related}
In this section, we introduce the background and related work of our hybrid sensing system. This section is divided into four subsections. The first two subsections provide an overview of the existing smart parking systems and security issues in IoT-sensing. Subsection \ref{ssec:crowdrelated} surveys crowdsensing and its security issues. Finally, Subsection \ref{ssec:hybridrelated} addresses the novelty of this work with a comparison with the existing literature. 

\subsection{Smart Cities and Smart Parking} \label{ssec:smartparking}
Smart cities and smart appliances \cite{aftabtiot} enrich our daily lives by providing information-driven services and resource management. Since the states of services and resources are dynamic and uncertain, smart cities often require real-time sensing and tracking. 

Among the service sectors in smart cities, many projects have developed vision-based smart parking systems with the development of computer vision technologies. Compared with traditional sensor-based approaches (e.g., using infrared \cite{infrared} or ultrasonic \cite{ultrasonic} sensors), vision-based systems have drastic improvements in detection accuracy \cite{approach}. Moreover, camera lenses are inexpensive and easy to install. Therefore, cameras can be integrated into embedded systems and then connected to the Internet to monitor parking lots in real-time.  

Note that most current studies focused on detecting the parking spaces in a parking lot. For example, \cite{vision-based} introduced a monitoring system where four cameras were set up in buildings around a parking lot, and the system can send and display real-time parking information. Similarly, \cite{real-time} proposed a system that counts the number of cars parked and identifies the vacant spaces. Furthermore, \cite{automated} suggested a method based on the coordinates of parking spaces. Compared to the methods above, \cite{deeplearning} presented a more robust solution using deep learning. Their system can infer the availability information of a large parking lot within a second. In this paper, we applied a machine learning method to infer the real-time occupancy information so that our system can provide a high detection accuracy and be integrated with crowdsourcing.

\subsection{IoT-Sensing} \label{ssec:iotsensing}
As \cite{securitysurvey} suggested that security and privacy issues tend to be ignored or have not been thoroughly investigated due to the development complexity of integrating both hardware and software components in most IoT applications. In the literature, the typical solutions are data perturbation and cryptography. For example, \cite{lightweight} implemented a privacy-preserving collaborative learning model, using a multiplicative random projection approach to obfuscate training data at each resource-limited IoT object. The authors claimed that their approach outperformed the other approaches based on additive noisification. Moreover, \cite{ppIoT} proposed a variational autoencoder which can achieve adversarial model-free anonymization. The results demonstrated the feasibility of anonymizing data in real-time on resource-constrained edge devices. Additionally, \cite{targetfinder} presented a privacy-preserving IoT-camera-based target locating system based on homomorphic encryption, which can protect the target's privacy and the video footages provided by IoT cameras. On the other hand, a less common yet powerful and straightforward solution is based on hardware. Hardware-based privacy-preserving methods have an innate advantage in security. Notably, \cite{optics} presented a privacy-preserving optical filter that can defocus the images before capture. However, their setup was complicated and thus difficult to deploy on a large scale. Besides, \cite{automatic} proposed a real-time occupancy recognition system for buildings using a low-cost embedded system. The authors put a frosted lens in front of the camera so that  people's identities in the video feed were no longer recognizable, but their experiments did not consider the effects of different blurriness levels.  

In parking monitoring applications, only a few papers have considered privacy when designing their systems. For instance, \cite{IOT-based} proposed a real-time sensor-based parking monitoring and billing system. The authors suggested that to protect user privacy, the sensors deployed should only broadcast sensitive information, such as the vehicle position, when a vehicle is inside the detection area of a parking lot. Moreover, \cite{wirelesssmart} implemented a smart camera network for parking monitoring, where the authors recommended executing all processing in cameras so that no information is sent out. However, both methods above are still vulnerable to attacks. Inspired by  \cite{automatic}, we implemented a hardware-based privacy-preserving IoT-sensing system by detecting vehicle flows in our preliminary work \cite{camera-based}. Based on our previous papers \cite{camera-based, previous}, we managed to further improve the method of detection by directly monitoring parking spaces.

\subsection{Crowdsensing} \label{ssec:crowdrelated}
Crowdsensing can enable smart cities to adapt to dynamic user behaviors and has been extensively applied to various systems in industries \cite{CFKF19surveyMCS}. Notable commercial platforms, like navigation apps (e.g., Waze, Google Maps, and Apple Maps), often collect users' mobility data (with or without users' awareness) to generate global knowledge of traffic conditions and patterns. Users can share various personal observations of services, such as bus arrival times, available parking spaces, and congestion statuses, to help other users plan their trips and activities \cite{parkingcrowdsensing, roadconditionsensing}. There are several projects (e.g., \cite{cmtseng2017dte,ckt2017phevopt}) that incorporate crowdsensing data into smart cities for the purposes of monitoring citizen mobility and managing transportation services. 

As data sharing is becoming more prevalent in smart cities, there have been substantial concerns over user privacy. In particular, any data leakage may lead to severe identity exposure since a large number of shared data contain sensitive information such as locations and faces. Therefore, a comprehensive solution is needed to protect users. Similar to IoT-sensing, two main approaches to preserve users' privacy in crowdsensing are data perturbation and cryptography. Among the data perturbation techniques, \cite{ppcrowdsensing22, ppcrowdsensing14} leveraged k-anonymity to cluster users into groups for anonymity preserving, whereas \cite{ppcrowdsensing2,ppcrowdsensing13,ppcrowdsensing16} utilized differential privacy to obfuscate sensing data and results. However, several studies (e.g.,\cite{deanony07,deanonymization2,deanonymization3}) have applied various techniques to show the practicality of deanonymizing users' hidden identities and other sensitive information based on their trajectories and other online activities. More importantly, since most data perturbation techniques introduce noises to the dataset, the data utility is reduced.
In contrast with data perturbation, other research studies have applied cryptographic techniques (e.g., homomorphic encryption and public-private key signatures \cite{ppcrowdsensing1,ppcrowdsensing25,ppcrowdsensing21,ppcrowdsensingzkp1}) to protect users' privacy. Among all the cryptographic techniques, zero-knowledge proofs has been widely applied in areas such as blockchain (e.g., \cite{blockchain1,blockchain2,blockchain3}), secure multi-party computation (e.g., \cite{MPC1,MPC2}), and electronic voting or auctions (e.g., \cite{voting1,voting2,auction1}). In our hybrid sensing system, we applied efficient zero-knowledge proofs, and designed privacy-preserving protocols that ensure user privacy in their participation.

\subsection{Comparisons to Our Work} \label{ssec:hybridrelated}
There are previous studies that hybridized crowdsensing with sensor networks in the areas such as pollen detection \cite{pollen}, indoor positioning \cite{positioning}, and urban data collection \cite{collaborative-data-collection}. However, to the best of our knowledge, this work is the first to integrate IoT-sensing and crowdsensing in parking space monitoring while considering privacy issues for both methods. Particularly, the comparisons between our privacy-preserving hybrid sensing system with the aforementioned papers are tabulated in Table \ref{tab:comparison}. 

\begin{table}[ht]
    \centering
    \caption{Comparisons between our privacy-preserving hybrid sensing system and previous studies.}
    {\scriptsize
    \begin{tabular}{@{}c@{}|c@{}c@{}c@{}|c@{}c@{}}
    \hline \hline
     & Crowd- & Privacy & Incentivization & IoT/Camera  & Privacy  \\ 
     & sensing &  Protection & & Sensing & Protection \\     
    \hline
     \cite{automatic,infrared,ultrasonic,approach,vision-based,real-time,automated,deeplearning} & $\times$ & $\times$  & $\times$ & $\checkmark$ & $\times$ \\
    \cite{targetfinder} & $\times$ & $\times$  & $\times$ & $\checkmark$ & Cryptographic \\
    \cite{lightweight,ppIoT} & $\times$ & $\times$ & $\times$ & $\checkmark$ & Data Perturbation \\
    \cite{optics} & $\times$ & $\times$ & $\times$ & $\checkmark$ & Optical Hardware \\
    \cite{IOT-based,wirelesssmart} & $\times$ & $\times$ & $\times$ & \checkmark & Software based \\
    \cite{camera-based} & $\times$ & $\times$ & $\times$ & $\checkmark$ & Physical Filter \\
    \cite{parkingcrowdsensing} & $\checkmark$ & $\times$  & $\checkmark$ & $\times$ & $\times$ \\
    \cite{roadconditionsensing,cmtseng2017dte,ckt2017phevopt} & $\checkmark$ & $\times$  & $\times$ & $\times$ & $\times$ \\
    \cite{ppcrowdsensing2,ppcrowdsensing22,ppcrowdsensing13,ppcrowdsensing14} & $\checkmark$ & Data Perturbation & $\checkmark$ & $\times$ & $\times$ \\
    \cite{ppcrowdsensing16} & $\checkmark$ & Data Perturbation & $\times$ & $\times$ & $\times$ \\
    \cite{ppcrowdsensing1,ppcrowdsensing25,ppcrowdsensing21,ppcrowdsensingzkp1} & $\checkmark$ & Cryptographic & $\checkmark$ & $\times$ & $\times$ \\
    \cite{ppcrowdsensingzkp2} & $\checkmark$ & Cryptographic & $\times$ & $\times$ & $\times$ \\
    \cite{positioning,pollen} & $\checkmark$ & $\times$  & $\times$ & $\checkmark$ & $\times$ \\
    \cite{collaborative-data-collection} & $\checkmark$ & $\times$  & $\checkmark$ & $\checkmark$ & $\times$ \\
    This work & $\checkmark$ & Cryptographic & $\checkmark$ & $\checkmark$ & Physical Filter \\
    \hline \hline
    \end{tabular}
    \label{tab:comparison}
    }
\end{table}

We highlight the novelty and key contributions of this paper as follows:
\begin{enumerate}
    \item We implemented a privacy-preserving hybrid sensing system based on parking monitoring using cameras  obscured by physical lenses at different blurriness levels.
    \item We applied a machine learning method to predict the occupancy of parking lots whose real-time video feed is blurred. The training process can be integrated with crowdsourcing, which improves detection accuracy while incentivizing voluntary users.
    \item We developed a mobile application for privacy-preserving crowdsensing parking monitoring using cryptographic techniques, which can be integrated with an IoT-sensing system.
\end{enumerate}

\section{Privacy-Preserving IoT-Sensing} \label{sec:IoT sensing}
\begin{figure*}[ht!]
 \centering
     \includegraphics[width=.6\textwidth]{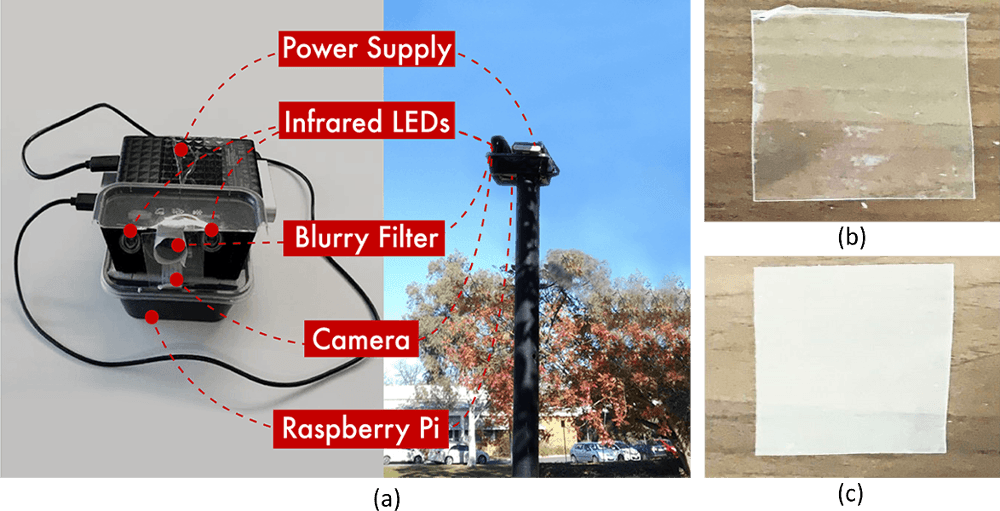}
       \caption{Hardware setup. (a) The Raspberry Pi camera system. (b) A blurry filter made of frosted glass. (c) A blurry filter made of frosted glass with translucent tape on both sides.}
      \label{fig:hardware}
\end{figure*}

In this section, we present the IoT-sensing part of our privacy-preserving hybrid sensing system for parking monitoring. To capture visual data from parking spaces, we deployed Raspberry Pi-based systems equipped with 160$^\circ$ fish-eye cameras, which can offer a wide panoramic field of view. The video streams captured have a resolution up to $1280\times720$, with a frame rate of 25 frames per second. The hardware setup is depicted in Fig.~\ref{fig:hardware} (a). Each device incorporates a blurry filter, a camera, two infrared LEDs, wireless connectivity, and a battery-based power supply. 

\subsection{Obscured Cameras with Blurry Filters}
In our camera system, the video footages are instantly processed on Raspberry Pi for rapid object recognition and classification. However, the system is inevitably connected to a wireless network for data sharing and hence is susceptible to hacking and attacks. In particular, the captured footages of parking spaces could be leaked. Private information such as vehicle license plates and the faces of pedestrians may be disclosed for unknown criminal purposes. To mitigate the privacy concern of our camera system, we deliberately obscure the cameras by hardware-based filters. We put frosted and blurry filters in front of the cameras to prevent capturing detailed footages of vehicles and pedestrians. Note that hardware-based filters are more robust than any software-based privacy methods, such as software-controlled blurring/resolution reduction or data deletion by discarding footages in system memory. Since the whole camera system could be hijacked by hackers, and any software-based privacy methods would be disabled readily, hardware-based filters provide a stronger privacy guarantee and cannot be circumvented without physical access to the sites. 

\begin{figure*}[th!]
 \centering
      \subfigure[]{\includegraphics[width=.19\textwidth]{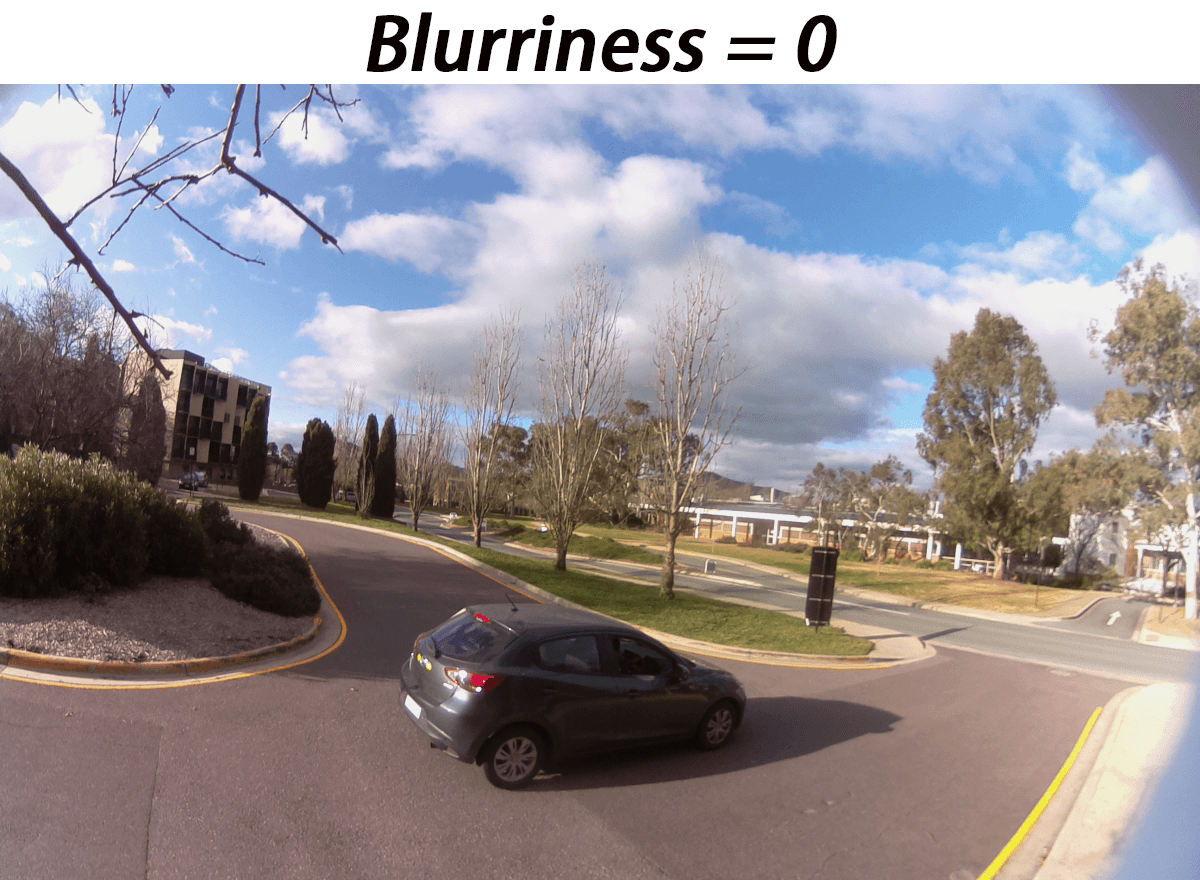}}
      \subfigure[]{\includegraphics[width=.19\textwidth]{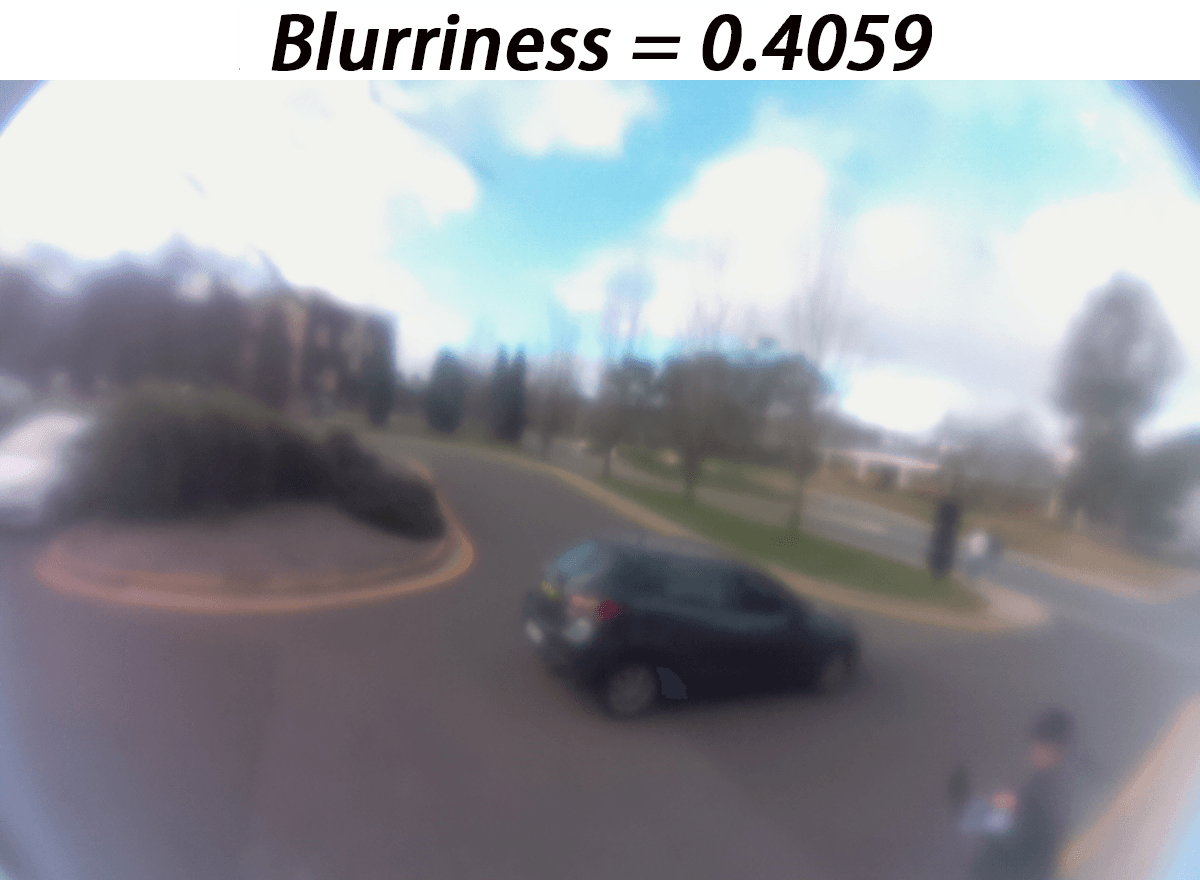}} \subfigure[]{\includegraphics[width=.19\textwidth]{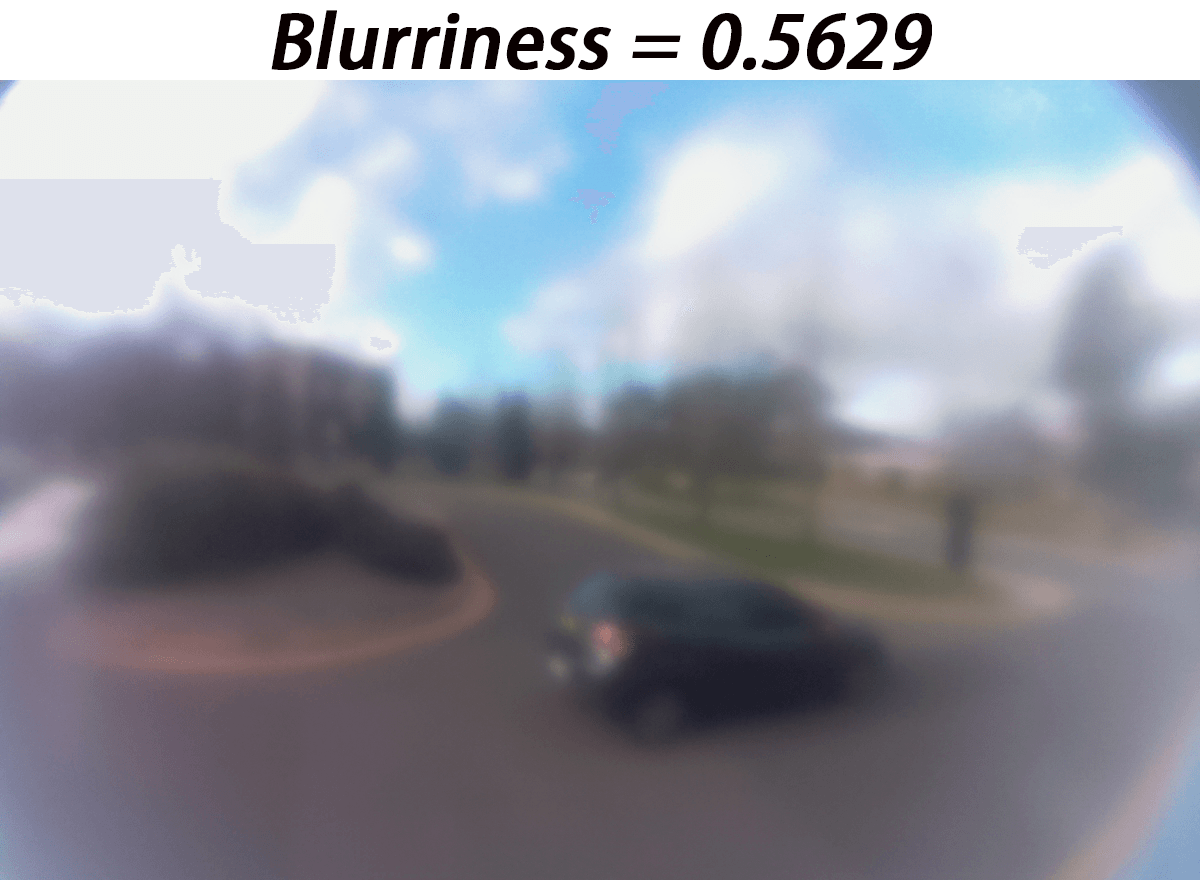}}
      \subfigure[]{\includegraphics[width=.19\textwidth]{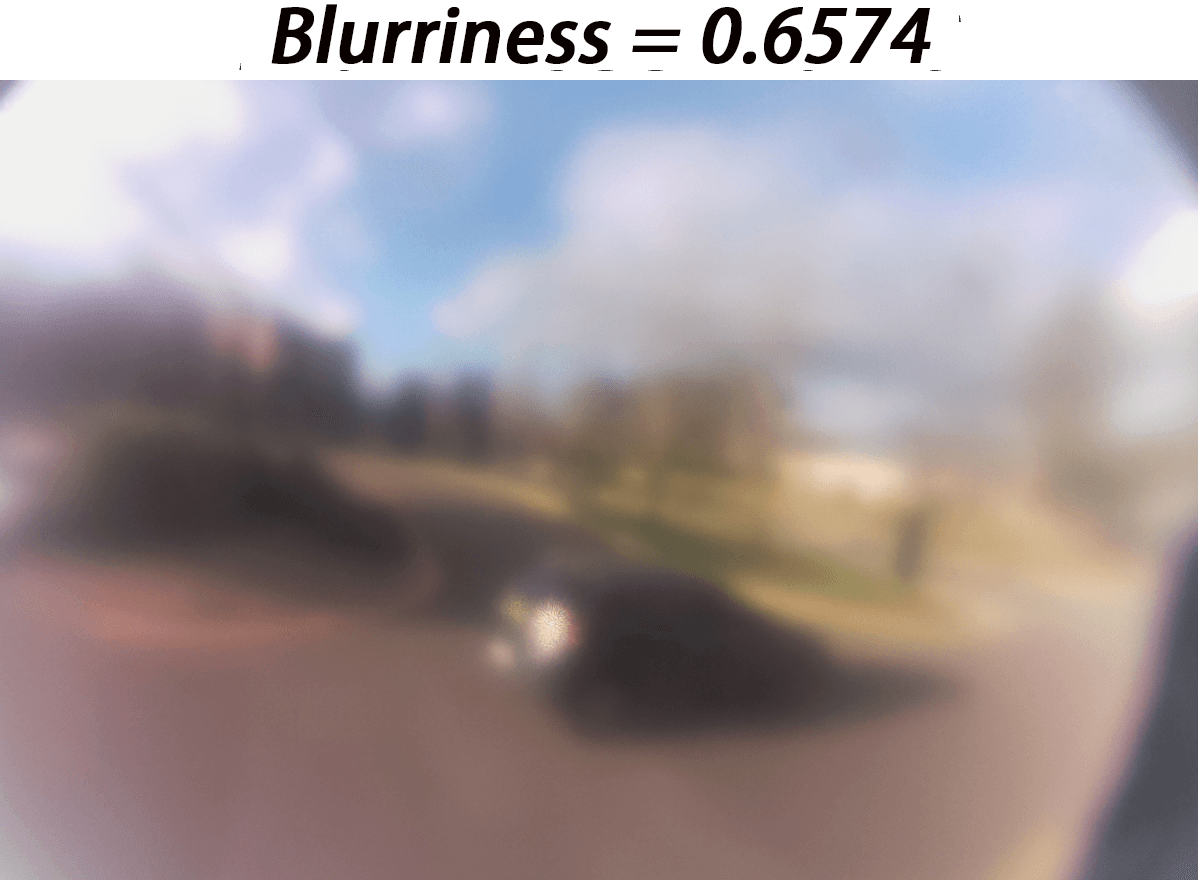}} \subfigure[]{\includegraphics[width=.19\textwidth]{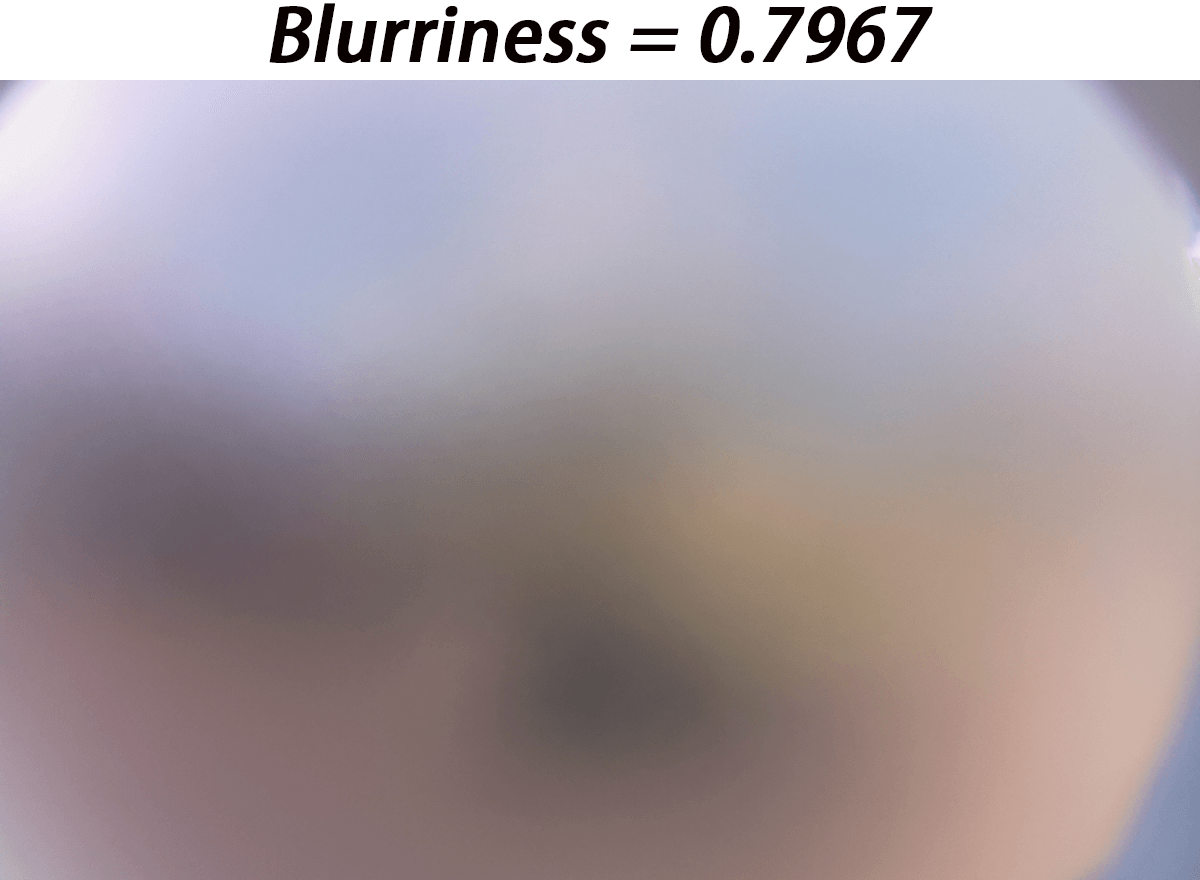}}
       \caption{Effects of using blurry filters. (a) The original image. (b) Using one frosted filter with translucent tape. (c) Using two frosted filters with translucent tape.
       (d) Using three frosted filters with translucent tape. (e) Using one frosted filter without translucent tape.}
      \label{fig:effects}
\end{figure*}

In this work, we applied various physical blurry filters to create different blurriness levels for the cameras. Fig.~\ref{fig:hardware} ($b$) and ($c$) show the frosted filters used in the system. Particularly, we adopted a piece of frosted glass in ($b$) and a piece of frosted glass with double-sided translucent tape in ($c$). The translucent tape can increase light transmittance and hence reduce blurriness, as the clearness of the filters shows. By obscuring one or more filters, we managed to simulate different levels of blurriness, as depicted in Fig.~\ref{fig:effects} ($a$), ($b$), ($c$), and ($d$). It is evident that as the blurriness level increases, it is increasingly hard to recognize any objects in the video. In this work, we consider the effectiveness of a blurred camera-based parking monitoring system. Such a system is considered privacy-preserving if sensitive information (such as the license plates of cars or the identity of the pedestrians) in the videos 
is not identifiable. According to the above definition, even the least blurred filter, as shown in Fig.~\ref{fig:effects} ($b$), satisfies our privacy preservation requirement. 

Next, we quantify the level of blurriness by a suitable metric for subsequent experimental evaluation. We applied a metric in \cite{edge} based on edge sharpness because it aligns well with human perception. This metric computes the average of the maximum difference of each pixel compared with its neighboring pixels. Given any blurred image and its original image, the blurriness level of the blurred image is defined as follows:    
    \begin{align}
         & {\tt Blurriness} =\frac{|X-Y|}{X} \label{ES} \\
         & {X=\frac{\sum_{i=1}^{M} \max_{j\in{\mathcal N}_i}(x_i-x_j)}{M}} \\
         & {Y=\frac{\sum_{i=1}^{N} \max_{j\in{\mathcal N}_i}(y_i-y_j)}{N}}  \\ \notag
    \end{align}
 where, $x_i$ and $y_i$ refer to the pixel values of the $i$-th pixel in the original and blurred images, respectively. $M$ and $N$ are the total numbers of pixels in the original and blurred images. ${\mathcal N}_i$ is the set of neighboring pixels of the $i$-th pixel. Based on the definition above, the blurriness levels are $0.4059$, $0.5629$, $0.6574$, and $0.7967$ for the various blurry filters in Fig.~\ref{fig:effects}. 

\subsection{Machine Learning for Parking Space Recognition under Blurry Filters}
\begin{figure}[th!]
  \centering
  \includegraphics[width=0.45\textwidth]{{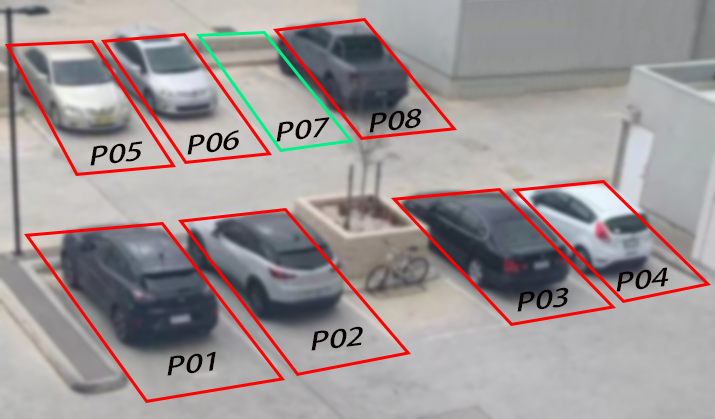}}
    \caption{An example of a captured parking space footage.}
    \label{fig:parking lot}
\end{figure}

After applying blurry filters, our system should still be capable of recognizing parking availabilities from camera-based sensors. This section describes the system that utilizes machine learning for parking space recognition executed on Raspberry Pi with blurry filters. Fig.~\ref{fig:parking lot} illustrates an example of our system capturing blurry footages of parking spaces via obscured cameras. The system administrators will first define a region of interest (namely, the parking spaces) on the screen.  
Through machine learning, the Raspberry Pi device can detect if there is a vehicle in each parking space (presented in different colors) based on the trained model. Note that the training data can be obtained through both IoT-sensing or crowdsourcing (e.g., Amazon Mechanical Turk).

\begin{figure}[th!]
  \centering
  \includegraphics[width=0.98\textwidth]{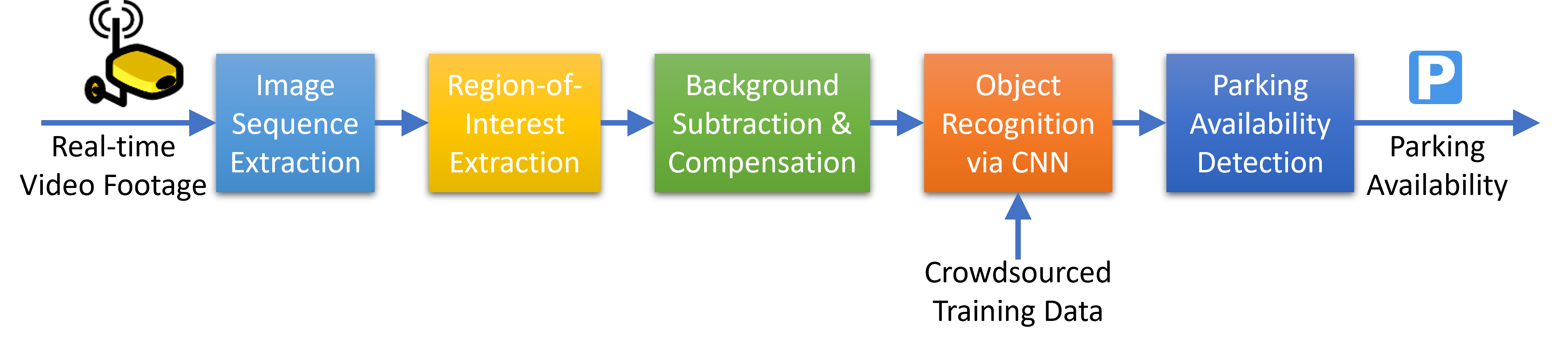}
    \caption{A flowchart showing the recognition algorithm.}
    \label{fig:flowchart}
\end{figure}

\begin{figure}[th!]
  \centering
  \includegraphics[width=0.5\textwidth]{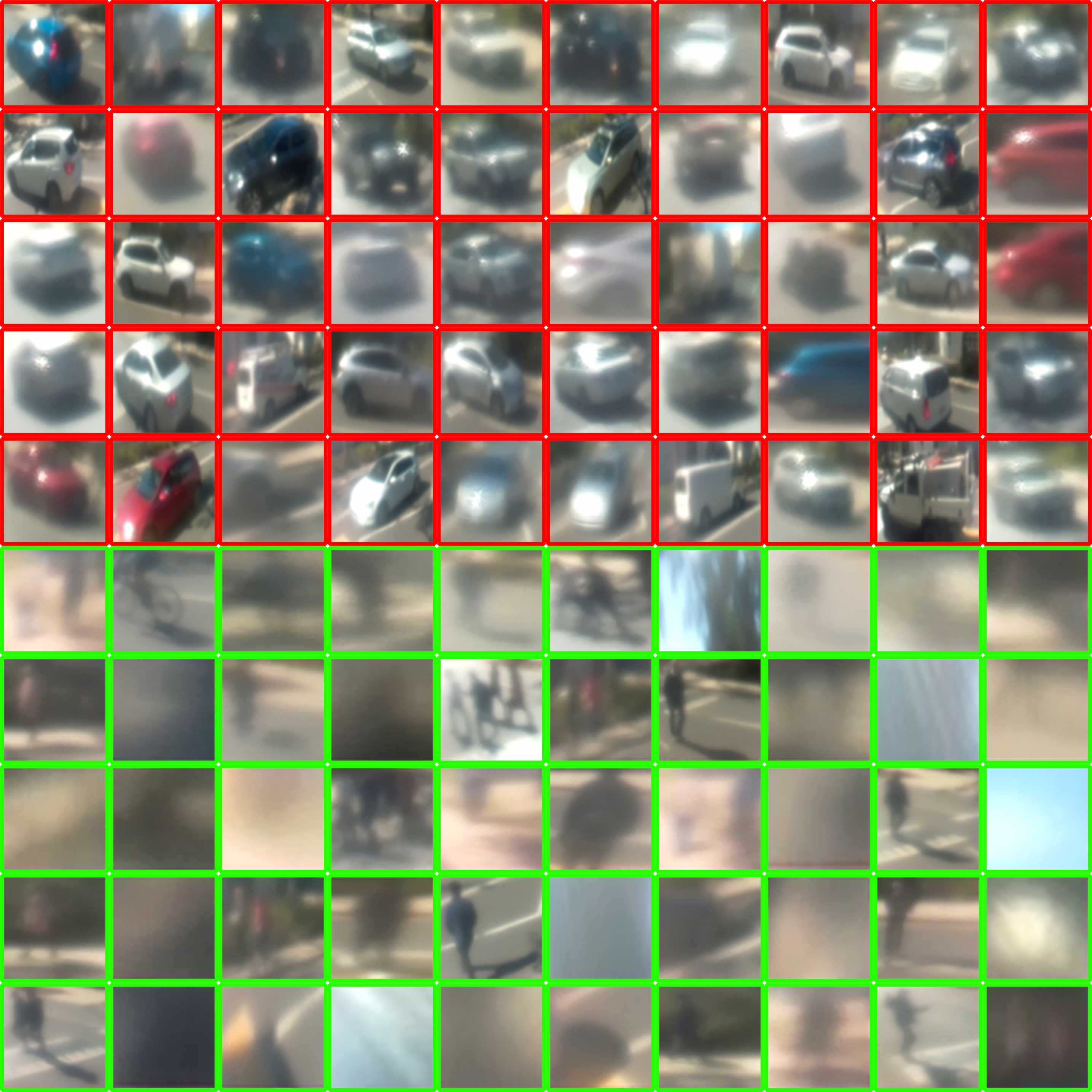}
  \caption{An example of extracted images with different blurriness levels used in the training sets where pictures with red frames are classified as cars, and pictures with green frames are non-cars.}
  \label{fig:training_data}
\end{figure}

\begin{figure*}[th!]
  \centering
  \includegraphics[width=\textwidth]{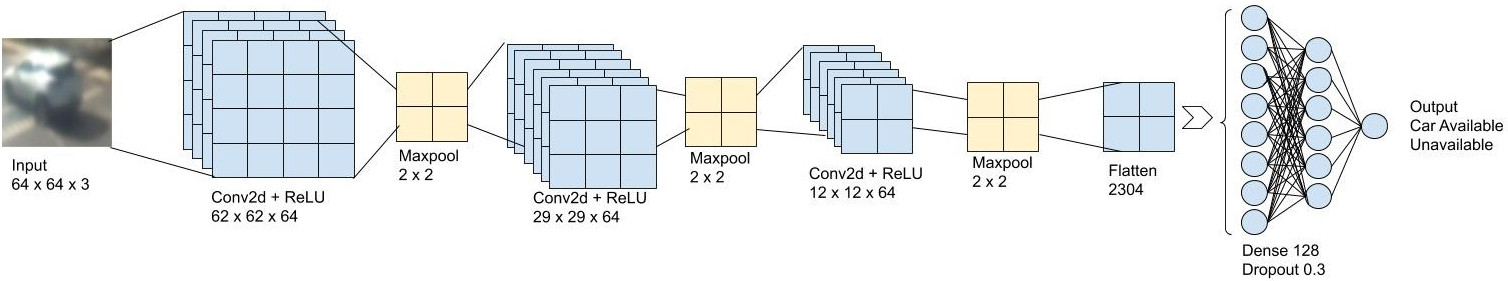}
  \caption{The CNN model used in this work.}
  \label{fig:CNN model}
\end{figure*}

A high-level flowchart of the recognition algorithm is depicted in Fig.~\ref{fig:flowchart}. The most critical components in the recognition algorithm are detecting the moving objects in the image as well as deciding if there are vehicles in the parking spaces. Specifically, the first method is implemented through background subtraction, where a background image is identified and compared to the current frame of the image. A moving object will be detected if the image is different from the background. For simplicity, we only consider large objects, such as vehicles and pedestrians, moving in the video. The detection will pause if all large objects in the video are observed to be stationary. 

To predict the availability of each parking space, we applied a Convolutional Neural Network (CNN) classifier, which is widely used in analyzing visual imagery with high accuracy. A typical architecture of CNN consists of an input layer, some convolutional layers, some pooling layers, a connected layer, and an output layer. An example of the training data is shown in Fig.~\ref{fig:training_data}, where some extracted images are classified as vehicles and non-vehicles. Practically, the cameras can be deployed almost anywhere to capture the training footages given that blurry filters are used. The CNN model will be more accurate and scalable by deploying a large scale of cameras in different locations so that the training set contains diverse images of vehicles pointed with different angles and heights, as illustrated in Fig.~\ref{fig:training_data}. Fig.~\ref{fig:CNN model} shows our CNN model, which includes three convolution blocks with a max pooling layer and a ReLU activation function in each of them. A fully connected layer, known as the dense layer, consists of 128 units on top of it and can be activated by a ReLU activation function and an L2 regularization. In our model, there is also a dropout layer to prevent overfitting. 

\subsection{Crowdsourcing Training Data for Machine Learning}
The training data for machine learning can be obtained through crowdsourcing. Specifically, the training datasets can be obtained from either an online crowdsourcing platform or the IoT sensors deployed in parking lots. The acquired datasets can effectively bootstrap the training process \footnote{In practice, the training process can be bootstrapped using online crowdsourced datasets if the on-site images captured from IoT sensors are insufficient.}, and the continuous data input from the IoT sensors can further improve detection accuracy. In our system, the IoT sensors extract some segmented images from the captured footages (e.g., Fig.~\ref{fig:training_data}), and users \footnote{We assume only registered users in our hybrid sensing system are allowed to contribute crowdsourced data generated from the IoT sensors because if those data are submitted to any crowdsourcing platform, malicious users could leverage all images available and potentially deanonymize our video streams.} will have the opportunities to contribute to the training sets by classifying whether there is a car or not in a segmented image (which is equivalent to generating labels) via a mobile app. Our system then applies the majority voting technique to aggregate the crowdsourced data and minimize errors in labels. Users will earn credits if they submit labels. More details are provided in Section~\ref{sec:crowdsensing} and Section~\ref{sec:implementation}. 
\section{Preliminaries of Cryptographic Components}  \label{sec:crypto}
Before presenting our privacy-preserving crowdsensing protocol, we introduce some relevant preliminaries of cryptographic components. Our privacy-preserving system relies on several basic cryptographic components, including cryptographic commitments, zero-knowledge proofs,  and public-private key signatures. More details can be found in a cryptography textbook \cite{crytobk}. 

First, we denote ${\mathbb Z}_p = \{0, ..., p-1\}$ by the set of integers modulo a large prime number $p$, which will be used for encrypting private data. For brevity, we simply write ``$x+y$'' and ``$x\cdot y$'' for modular arithmetic without explicitly mentioning ``${\tt mod\ } p$''. We consider a finite group ${\mathbb G}$ of order $p$. Then, we pick $g, h$ as two generators of ${\mathbb G}$, such that they can generate every element in ${\mathbb G}$ by proper powers, namely, for each $e \in {\mathbb G}$, there exists $x,y \in {\mathbb Z}_p$ such that $e = g^x = h^y$. The classical discrete logarithm assumption states that given $g^x$, it is computationally hard to obtain $x$, which underlies the security of many cryptosystems, including ours.

\subsection{Cryptographic Commitments}
A cryptographic commitment allows a user to hide a secret (e.g., crowdsensed data). We use the Pedersen commitment, which is perfectly hiding (i.e., an adversarial cannot unlock the secret) and computationally binding (i.e., an adversarial cannot associate with another secret in polynomial time). To commit a secret value $x\in{\mathbb Z}_p$, a user first picks a random number ${\tt r} \in {\mathbb Z}_p$ to mask the commitment. Then, he computes the commitment by ${\tt Cm}(x, {\tt r}) = g^x \cdot h^{\tt r}$. Note that the Pedersen commitment satisfies the homomorphic property: ${\tt Cm}(x_1+x_2, {\tt r}_1+{\tt r}_2) = {\tt Cm}(x_1, {\tt r}_1) \cdot {\tt Cm}(x_2, {\tt r}_2)$. In this paper, we simply write ${\tt Cm}(x)$ without specifying ${\tt r}$. 

\subsection{Zero-knowledge Proofs}
In a zero-knowledge proof, a prover (e.g., crowdsensing user) convinces a verifier (e.g., server) of the knowledge of a secret without revealing the secret (e.g., crowdsensed data). For example, to show the knowledge of $(x, {\tt r})$ for ${\tt Cm}(x, {\tt r})$ without revealing $(x, {\tt r})$. A zero-knowledge proof should satisfy completeness (i.e., a prover knowing the secret can always convince the verifier), soundness (i.e., a prover not knowing the secret cannot convince the verifier), and zero-knowledge (i.e., the verifier cannot learn the secret).

\subsubsection{$\Sigma$-Protocol}
\
A general approach to provide zero-knowledge proofs is based on {\em $\Sigma$-protocol}. This can prove the knowledge of $x$ such that $f(x)=y$ with concealed $x$, where $y$ and $f(\cdot)$ are revealed and $f(\cdot)$ is not computationally invertible. Suppose $f(\cdot)$ satisfies the homomorphic property: $f(a+b) = f(a) + f(b)$. The $\Sigma$-protocol proceeds as follows:

\begin{enumerate}[leftmargin=*]
\item First, the prover sends a commitment $y' = f(x')$ for a random $x'$, to the verifier. 
\item Next, the verifier generates a random challenge $\beta$ and sends it to the prover. 
\item The prover replies with $z = x' + \beta \cdot x$ \mbox{(which does not reveal $x$)}. 
\item Finally, the verifier checks whether $f(z) \overset{?}{=} y' + \beta \cdot y$. 
\end{enumerate}

One can show that $\Sigma$-protocol satisfies completeness, soundness, and zero-knowledge. Next, we present some examples of $\Sigma$-protocol that will be utilized in our protocol.

\subsubsection{Zero-knowledge Proof of Committed Value ({\tt zkpCm})}
\
Given ${\tt Cm}(x, {\tt r})$, a prover wants to convince a verifier of the knowledge $(x, {\tt r})$. $\Sigma$-protocol can be applied as follows:

\begin{enumerate}[leftmargin=*]
\item The prover first randomly generates $(x', {\tt r}') \in {\mathbb Z}^2_p$ and sends the corresponding commitment ${\tt Cm}(x', {\tt r}')$ to the verifier.
\item The verifier sends a random challenge $\beta \in {\mathbb Z}_p$ to the prover. 
\item The prover replies with $z_x = x' + \beta \cdot x$ and $z_{\tt r} = {\tt r}' + \beta \cdot {\tt r}$.
\item The verifier checks whether $g^{z_x} \cdot h^{z_{\tt r}} \overset{?}{=}  {\tt Cm}(x', {\tt r}') \cdot {\tt Cm}(x, {\tt r})^\beta$.
\end{enumerate}

Denote a zero-knowledge proof of committed value for ${\tt Cm}(x, {\tt r})$ by ${\tt zkpCm}[x]$. If a prover wants to convince a verifier of the knowledge of $x$ in ${\tt Cm}(x, {\tt r})$ with a given random mask ${\tt r}$, the prover can set ${\tt r}' = 0$ and the verifier can construct $z_{\tt r} = \beta \cdot {\tt r}$ by himself.

\subsubsection{Zero-knowledge Proof of Membership ({\tt zkpMbs})}
\
Given a set ${\mathcal X} = \{x_1, ..., x_n \}$ and ${\tt Cm}(x, {\tt r})$, a prover wants to convince a verifier of the knowledge $x \in {\mathcal X}$ without revealing $x$. $\Sigma$-protocol can be applied as follows:

\begin{enumerate}[leftmargin=*]
\item Suppose $x=x_i \in {\mathcal X}$. The prover first randomly generates $(x'_j, {\tt r}'_j) \in {\mathbb Z}_p^2$ and computes the commitment ${\tt Cm}(x'_j, {\tt r}'_j)$ for all $j\in\{1,...,n\}$. Then, the prover randomly generates $\beta_j\in {\mathbb Z}_p$ for each $j\in\{1,...,n\}\backslash\{i\}$, and computes
\[
z_{x_j} = 
\begin{cases}
x'_j + (x_i - x_j) \beta_j,& \mbox{if\ } j\in\{1,...,n\}\backslash\{i\}\\
x'_i, & \mbox{if\ } j = i\\
\end{cases}
\] 
Next, the prover sends $({\tt Cm}(x'_j, {\tt r}'_j), z_{x_j})_{j=1}^n$ to the verifier.
\item The verifier sends a random challenge $\beta\in {\mathbb Z}_p$ to the prover. 
\item The prover sets $\beta_i = \beta - \sum_{j\ne i}\beta_j$, then computes $z_{{\tt r}_j} = {\tt r}'_j + {\tt r} \cdot \beta_j$ for all $j\in\{1,...,n\}$, and sends $(\beta_j, z_{{\tt r}_j})_{j=1}^n$ to the verifier.
\item The verifier checks whether $\beta \overset{?}{=} \sum_{i=1}^n \beta_j$ and 
\[
g^{z_{x_j}} \cdot h^{z_{{\tt r}_j}} \overset{?}{=} {\tt Cm}(x'_j, {\tt r}'_j) \cdot \Big( \frac{{\tt Cm}(x, {\tt r})}{g^{x_j}} \Big)^{\beta_j} \mbox{for all\ } j\in\{1,...,n\}
\]
\end{enumerate}

Denote a zero-knowledge proof of member for $x \in {\mathcal X}$ by ${\tt zkpMbs}[x, {\mathcal X}]$.

\subsubsection{Zero-knowledge Proof of Non-negativity ({\tt zkpNN})}
\
Given ${\tt Cm}(x, {\tt r})$, a prover wants to convince a verifier of the knowledge of $x \ge 0$ without revealing $x$. Suppose $x < 2^m$. We aim to prove there exist $(b_1, ..., b_m)$ such that $b_i \in \{0, 1\}$ for $i \in \{0, ..., m\}$ and $\sum_{i=1}^{m}b_i \cdot 2^{i-1} = x$. $\Sigma$-protocol can be applied as follows:

\begin{enumerate}[leftmargin=*]
\item The prover sends $({\tt Cm}(b_i, {\tt r}_i))_{i=1}^{m}$ to the verifier, and provides ${\tt zkpMbs}[b_i, \{0,1\}]$ for each $b_i$ to prove that $b_i \in \{0, 1\}$. Also, the prover randomly generates ${\tt r}'\in {\mathbb Z}_p$ and sends the commitment ${\tt Cm}(0, {\tt r}')$ to the verifier.
\item The verifier sends a random challenge $\beta\in {\mathbb Z}_p$ to the prover. 
\item The prover replies with $z_{\tt r} = {\tt r}' + \beta \cdot (\sum_{i=1}^m {\tt r}_i \cdot 2^{i-1} - {\tt r})$.
\item The verifier checks whether $h^{z_{\tt r}} \overset{?}{=}  {\tt Cm}(0,{\tt r}') \cdot {\tt Cm}(x, {\tt r})^{-\beta}\cdot\prod_{i=1}^m {\tt Cm}(b_i, {\tt r}_i)^{\beta\cdot 2^{i-1}}$
\end{enumerate}

Denote a zero-knowledge proof of the non-negativity of $x$ by ${\tt zkpNN}[x]$.

\subsection{Non-interactive Zero-knowledge Proofs}
The interactive $\Sigma$-protocol can be converted to a non-interactive zero-knowledge proof by Fiat-Shamir heuristics which remove the challenge provided by the verifier. Let ${\mathcal H}(\cdot)\mapsto {\mathbb Z}_p$ be a cryptographic hash function. Given a list of commitments (${\tt Cm}_1, ..., {\tt Cm}_r$), one can map to a single hash value by  ${\mathcal H}({\tt Cm}_1|...|{\tt Cm}_r)$, where the input is the concatenated string of (${\tt Cm}_1, ..., {\tt Cm}_r$). In a $\Sigma$-protocol, one can set the challenge $\beta = {\mathcal H}({\tt Cm}_1|...|{\tt Cm}_r)$, where (${\tt Cm}_1, ..., {\tt Cm}_r$) are all the commitments generated by the prover prior to the step of verifier-provided challenge (Step 2 of $\Sigma$-protocol). Hence, the prover does not need to wait for the challenge from the server but instead generates the random challenge himself. The verifier will generate the same challenge with the same procedure for verification. We denote the non-interactive versions of the previous zero-knowledge proofs by {\tt nzkpCm}, {\tt nzkpSum}, {\tt nzkpMbs}, and {\tt nzkpNN}, respectively. 

\subsection{Public-Private Key Signatures}
Cryptographic signatures can verify the authenticity of some given data. Suppose a signer has a pair of public and private keys $(K^{\tt p}, K^{\tt s})$ for an asymmetric key cryptosystem (e.g., RSA). To sign a message $m$, the signer first maps $m$ by a cryptographic hash function ${\mathcal H}(m)$ (e.g., SHA-3). Then, the signature of $m$ is the encryption ${\tt sign}_{K^{\tt s}}(m) = {\tt Enc}_{K^{\tt s}}[{\mathcal H}(m)]$. Given $(m, K^{\tt p})$, anyone can verify the signature ${\tt sign}_{K^{\tt s}}(m)$ by checking whether the decryption ${\tt Dec}_{K^{\tt p}}[{\tt sign}_{K^{\tt s}}(m)]  \overset{?}{=} {\mathcal H}(m)$.

\section{Privacy-Preserving Crowdsensing} \label{sec:crowdsensing}
In this section, we present the crowdsensing part with a privacy-preserving crowdsensing protocol, which is complementary to the privacy-preserving IoT-sensing system presented in Section~\ref{sec:IoT sensing}. The protocol draws on the basic cryptographic components of cryptographic commitments, zero-knowledge proofs, and public-private key signatures.

\subsection{Threat Model} \label{sec:threat model}
Our crowdsensing system aims to preserve the privacy of contributing users while being robust against the possible attacks of dishonest users. Firstly, dishonest users may submit multiple (sometimes inconsistent) crowdsensed data entries of a single parking space to the server. Since our protocol is privacy-preserving, it should not reveal users' identities, which enables dishonest users to potentially earn more credits. Hence, the submitted crowdsensed data entries should be verified to prevent duplication and inconsistency while preserving user anonymity. Secondly, dishonest users may attempt to claim more credits for their contributed data than they ought to. We need to ensure the consistency of credit claiming with respect to users' actual contributions, subject to anonymity. In particular, our protocol supports the following requirements of privacy-preserving crowdsensing:

\begin{enumerate}[leftmargin=*]
\item {\bf Anonymity}: The identities of users should not be revealed or leaked from the server data. However, all data may be associated with only randomized identification. 
\item {\bf Unlinkability}: Each data entry cannot be linked to reveal the data sources. One cannot distinguish whether the data entries are from different users or the same user.
\item {\bf Cheating Prevention}: The protocol should be able to prevent cheating behaviors (e.g., duplicate data submissions, inconsistent credit claims), even though the data are anonymized and unlinkable to a specific user. 
\item {\bf Incentive Support}: Users can still earn appropriate credits with verification for their useful data contributions in a privacy-preserving manner.
\end{enumerate}

Note that there are several assumptions in our model. We only consider anonymity in the server, concerning the recorded persistent data, not in the communication channels. Also, we assume that the communication channels are protected from external attacks, such as eavesdropping, IP hijacking, and  man-in-the-middle attack. Finally, users can use proper security communication systems to ensure anonymity and security in the communication channels, such as anonymous routing and IPSec. 

\begin{table}[t!]  
\caption{Notations used in the protocol.} \label{tab:notations}
\begin{tabular}{@{}c@{}|p{0.85\textwidth}@{}}
\hline \hline
$K^{\tt p}$  & Public key of the server          \\
$K^{\tt s}$   &   Private key of the server \\
${\tt Cm}(x, {\sf r})$ & Cryptographic commitment of $x$ and random mask ${\sf r}$\\
$Q$ & Credential of a user\\
${\tt sign}_Q$ & Server's signature on the credential $Q$\\
$s$ & User's secret key\\
$q$ & Credential identifier\\
$b$ & User's balance of credits\\
$b_0$ & User's default initial balance\\
$c_{\tt q}$ & Credits required for each parking availability inquiry\\
$R$ & Crowdsensed data entry of a user for the $j$-th parking space at time $t$  \\
$\tau^{j,t}$ & User's ticket associated with each data entry\\ 
$a^{j,t}$ & User's indicator of the observed availability of the $j$-th parking space at time $t$\\
$v^{j,t}$ & Aggregate outcome of the submitted data entries $\{a^{j,t}\}$ \\
$c^{j,t}$ & Eligible credit of the data entry for ticket $\tau^{j,t}$\\
$\mathbb{T}_{\tt D}$ & Table of crowdsensed data entries\\
$\mathbb{T}_{\tt C}$ & Table of eligible credits\\
$\mathbb{T}_{\tt Q}$ & Table of used credentials\\
\hline \hline
\end{tabular}
\end{table}

\subsection{Privacy-Preserving Protocol}
\begin{figure}[ht]
\centering
\begin{minipage}{0.88\textwidth}
    \centering
    \subfigure[Registration]{\includegraphics[width=0.4\textwidth]{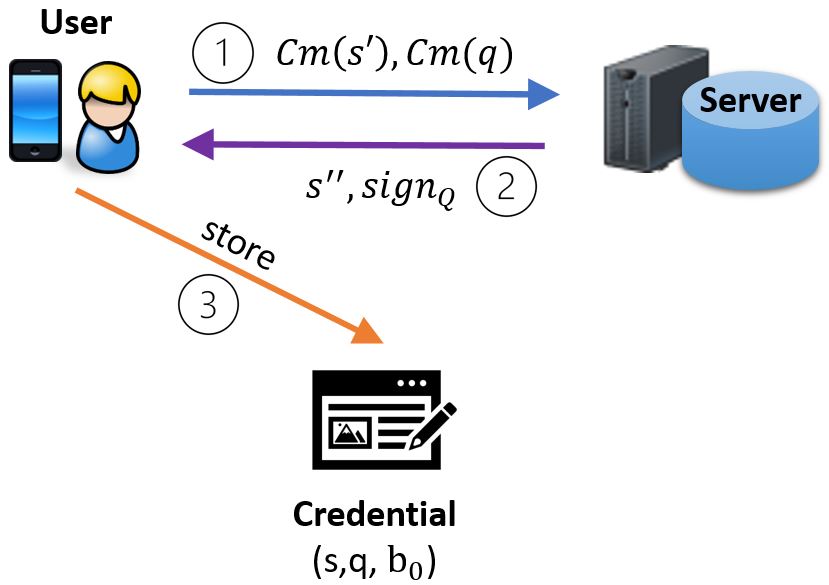}}  \qquad 
    \subfigure[Data submission]{\includegraphics[width=0.5\textwidth]{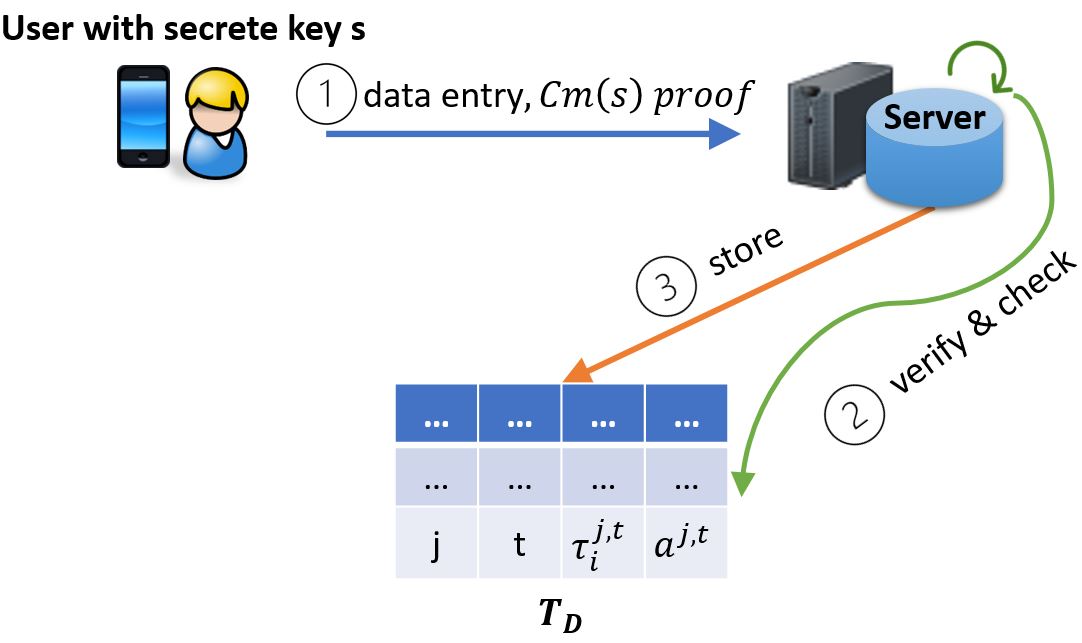}}\qquad
     \medskip
    \subfigure[Credit claiming]{\includegraphics[width=0.5\textwidth]{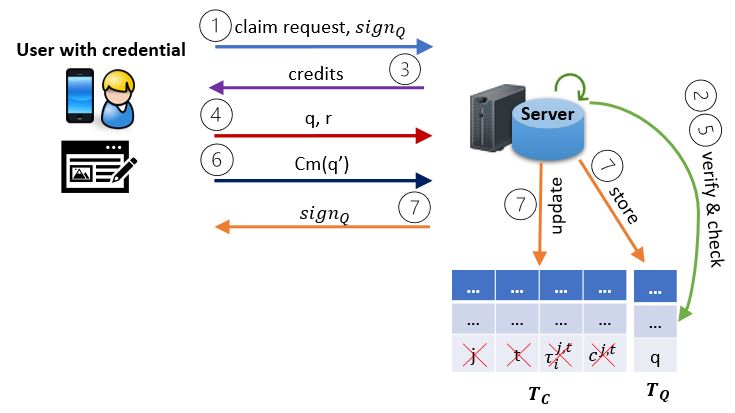}} 
    \subfigure[Result inquiry]{\includegraphics[width=0.46\textwidth]{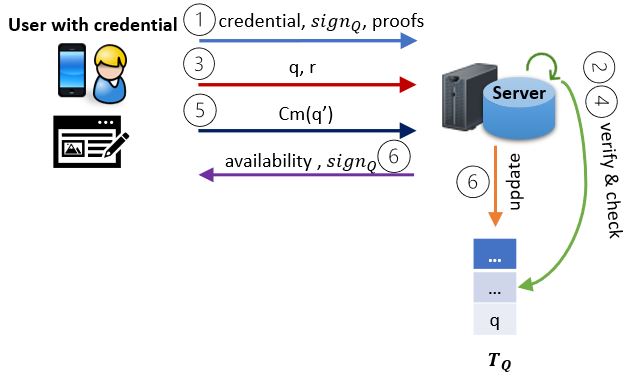}}\qquad
 \end{minipage}
     \caption{An illustration of the key stages in the privacy-preserving crowdsensing protocol. (a) The registration stage. (b) The data submission stage. (c) The credit claiming stage. (d) The result inquiry stage.}
    \label{fig:protocol steps}
\end{figure}

In this subsection, we present the details of our privacy-preserving protocol between the server and users. Our protocol ensures anonymity, unlinkability, cheating prevention, and incentive support. Some key notations are listed in Table~\ref{tab:notations}. There are five stages in our protocol, which are system setup, user registration, crowdsensed data submission, credit claiming, and result inquiry. See Fig.~\ref{fig:protocol steps} for an illustration of the key stages in the privacy-preserving crowdsensing protocol.

\subsubsection{System Setup:} \

First, the server generates the cryptographic parameters and notifies all users. The server chooses a proper ${\mathbb Z}_p$ and a finite group ${\mathbb G}$ of order $p$ with generators $g, h$. The server also sets an appropriate collision-resistant cryptographic hash function ${\mathcal H}(\cdot)$. The cryptographic parameters should be decided to meet the desired security and efficiency requirements.

\subsubsection{User Registration:} \

When a new user joins the crowdsensing system, the user needs to register with the server. After registration, the user obtains a credential $Q = (s, q, b)$, which consists of a secret key $s$ (used to claim credits), a credential identifier $q$ (used to verify the freshness of the credential and prevent double-spending), and a balance of credits $b$. The steps of this stage are illustrated in Fig.~\ref{fig:protocol steps} (a). The details of the steps are described as follows.

\smallskip

\begin{longfbox}[border-break-style=none,border-color=\#bbbbbb,background-color=\#eeeeee,breakable=true,width=0.98\textwidth]
{\em Registration Stage Details}: 

\begin{enumerate}[leftmargin=*]

 \item A new user randomly generates $(s', q) \in \mathbb{Z}^2_p$, and then sends cryptographic commitments ${\tt Cm}(s')$ and ${\tt Cm}(q)$ to the server.
 \item The server randomly picks $s'' \in \mathbb{Z}_p$, and computes ${\tt Cm}(s) = {\tt Cm}(s')\cdot{\tt Cm}(s'',0)$. Then, the server generates a signature ${\tt sign}_Q = {\tt sign}({\tt Cm}(s)|{\tt Cm}(q)|{\tt Cm}(b_0))$, where $b_0$ is the default initial balance. Next, the server sends $(s'', {\tt sign}_Q)$ to the user.
 \item The user computes $s=s'+s''$, and stores $Q=(s, q, b_0)$ with ${\tt sign}_Q$. Here, $s$ and $q$ are not known by the server.  
 
\end{enumerate}
\end{longfbox}

\smallskip
Note that secret key $s$ is jointly generated by a user and the server to prevent replay attack since any user may reuse another user's ${\tt Cm}(s)$ and ${\tt Cm}(q)$ in the registration to masquerade another user. 

\subsubsection{Crowdsensed Data Submission:}\ 

After the registration, the user can submit crowdsensed data (i.e., observed parking availabilities) to the server. A crowdsensed data entry will be associated with a secret key $s$ in a privacy-preserving manner. The user can claim credits for verified useful contributions subsequently by $s$. Let $a^{j,t}$ be a user's observed indicator variable of the availability of the $j$-th parking space at time $t$. $a^{j,t} = 1$ indicates that the parking space is available. Otherwise, $a^{j,t} = 0$. The steps of this stage are illustrated in Fig.~\ref{fig:protocol steps} (b). The details of the steps are described as follows.

\smallskip

\begin{longfbox}[border-break-style=none,border-color=\#bbbbbb,background-color=\#eeeeee,breakable=true,width=0.98\textwidth]
{\em Data Submission Stage Details}: 
\begin{enumerate}[leftmargin=*]
    \item The user computes $\tau^{j,t} ={\tt Cm}(s, {\mathcal H}(j|t))$. The user also generates ${\tt nzkpCm}[s]$ for ${\tt Cm}(s, {\mathcal H}(j|t))$ with a known random mask ${\mathcal H}(j|t)$.		
	\item The user sends the data entry $R = (j, t, \tau^{j,t}, a^{j,t})$ along with ${\tt nzkpCm}[s]$ to the server.
	\item The server verifies ${\tt nzkpCm}[s]$, and checks if there is a duplicate data entry with the same $(j, t, \tau_i^{j,t})$ on the table of crowdsensed data entries $\mathbb{T}_{\tt D}$. The server also checks time $t$.  $R$ will be rejected, if ${\tt nzkpCm}[s]$ fails, or there is a duplicate entry on $\mathbb{T}_{\tt D}$, or $t$ is not the recent time slot.
\end{enumerate}
\end{longfbox}

\smallskip
Note that the ticket $\tau^{j,t}$ using ${\mathcal H}(j|t)$ for the random mask is to prevent a user from submitting duplicate data entries. The secret key $s$ in ${\tt Cm}(s, {\mathcal H}(j|t))$ ensures that the user is the only one who can claim the credit subsequently.

\subsubsection{Credit Claiming:}\

The users can claim credits from their verified contributions, which can be redeemed when inquiring  about the results of the crowdsensing system. The server will periodically update $v^{j,t}$, the aggregate outcome of the submitted data entries $\{a^{j,t'}\}$. A simple aggregation is based on the majority vote of $\{a^{j,t'} : |t-t'| \le \epsilon\}$. Namely, the server takes a majority vote on the set of submitted data entries within time interval $[t - \epsilon,  t + \epsilon]$. If the user's data entry matches the outcome of the majority vote of all users, i.e., $a^{j,t'} = v^{j,t}$ for $t' \in [t - \epsilon,  t + \epsilon]$, then the contributed users will be eligible for credits. 

Next, the server will add $C = (j, t, \tau^{j,t}, c^{j,t})$ to the table of eligible credits $\mathbb{T}_{\tt C}$, where $c^{j,t}$ is the credits for the data entry with corresponding ticket $\tau^{j,t}$. An updated credential with a new credential identifier and a new balance will be created. A user with ticket $\tau^{j,t}$ can claim the respective credits in the protocol. The steps of this stage are illustrated in Fig.~\ref{fig:protocol steps} (c). The details of the steps are described as follows.

\smallskip

\begin{longfbox}[border-break-style=none,border-color=\#bbbbbb,background-color=\#eeeeee,breakable=true,width=0.98\textwidth]
{\em Credit Claiming Stage Details}: 
\begin{enumerate}[leftmargin=*]
    \item The user submits claim request $(j, t, \tau^{j,t})$ to the server, along with credential commitments $({\tt Cm}(s), {\tt Cm}(q), {\tt Cm}(b))$, signature ${\tt sign}_Q$, and ${\tt nzkpCm}[s]$.
	\item The server verifies ${\tt sign}_Q$ and ${\tt nzkpCm}[s]$ based on $({\tt Cm}(s)$, ${\tt Cm}(q),$ ${\tt Cm}(b))$. The claim will be rejected, if ${\tt sign}_Q$ or ${\tt zkpCm}[s]$ fails.
	\item The server returns $C = (j, t, \tau^{j,t}, c^{j,t})$ for the corresponding ticket $\tau^{j,t}$ to the user.
	\item The user proceeds to update his credential by revealing $(q, {\tt r})$ in ${\tt Cm}(q)$ to the server.
	\item The server checks if $(q, {\tt r})$ are committed values of ${\tt Cm}(q)$, and check $q$ in the  table of used credentials $\mathbb{T}_{\tt Q}$. The claim will be rejected, if ${\tt Cm}(q)$ check fails or $q$ is already on $\mathbb{T}_{\tt Q}$. 
	\item The user randomly generates $q' \in \mathbb{Z}_p$ and sends ${\tt Cm}(q')$ to the server.
	\item The server first computes ${\tt Cm}(b') =  {\tt Cm}(b)\cdot{\tt Cm}(c^{j,t} ,0)$. Then, the server generates signature ${\tt sign}_Q = {\tt sign}({\tt Cm}(s)|{\tt Cm}(q')$ $|{\tt Cm}(b'))$, sends ${\tt sign}_Q$ to the user, adds $q$ to $\mathbb{T}_{\tt Q}$, and removes $C = (j, t, \tau^{j,t}, c^{j,t})$ from $\mathbb{T}_{\tt C}$.
\end{enumerate}
\end{longfbox}

\smallskip

Note that the server needs to update $\mathbb{T}_{\tt Q}$ and $\mathbb{T}_{\tt C}$ after each claim to prevent submitting duplicate claims.

\subsubsection{Result Inquiry:}\

A user can make inquiry about the results (i.e., the availability of parking spaces) by spending an amount of $c_{\tt q}$ credits. The user needs to prove that his balance in the credential is sufficient, namely, $b - c_{\tt q} \ge 0$, in a privacy-preserving manner. An updated credential with a new credential identifier and a new balance will be created. The steps of this stage are illustrated in Fig.~\ref{fig:protocol steps} (d). The details of the steps are described as follows. 

\smallskip

\begin{longfbox}[border-break-style=none,border-color=\#bbbbbb,background-color=\#eeeeee,breakable=true,width=0.98\textwidth]
{\em Result Inquiry Stage Details}: 
\begin{enumerate}[leftmargin=*]
    \item The user submits credential commitments $({\tt Cm}(s), {\tt Cm}(q)$, ${\tt Cm}(b))$, signature ${\tt sign}_Q$, ${\tt nzkpCm}[s]$, and ${\tt nzkpNN}[b - c_{\tt q}]$ to the server.
	\item The server checks ${\tt sign}_Q$, ${\tt nzkpCm}[s]$ and ${\tt nzkpNN}[b - c_{\tt q}]$. The inquiry will be rejected, if checking ${\tt sign}_Q$, ${\tt nzkpCm}[s]$ or ${\tt nzkpNN}[b - c_{\tt q}]$ fails.
	\item The user proceeds to update his credential by revealing $(q, {\tt r})$ in ${\tt Cm}(q)$ to the server.
	\item The server checks if $(q, {\tt r})$ are committed values of ${\tt Cm}(q)$, and checks $q$ in the  table of used credentials $\mathbb{T}_{\tt Q}$. The claim will be rejected, if ${\tt Cm}(q)$ check fails or $q$ is already on $\mathbb{T}_{\tt Q}$. 
	\item The user randomly generates $q' \in \mathbb{Z}_p$ and sends ${\tt Cm}(q')$ to the server.
	\item The server first returns the requested availability information. Next, the server computes ${\tt Cm}(b') =  {\tt Cm}(b)\cdot{\tt Cm}(- c_{\tt q} ,0)$, generates ${\tt sign}_Q = {\tt sign}({\tt Cm}(s)|{\tt Cm}(q')|{\tt Cm}(b'))$, sends ${\tt sign}_Q$ to the user, and adds $q$ to $\mathbb{T}_{\tt Q}$.
\end{enumerate}
\end{longfbox}

\medskip

Finally, our crowdsensing protocol can be shown to satisfy anonymity and unlinkability. 
\section{System Implementation}\label{sec:implementation}
In this section, we present the system implementation of the hybrid sensing system, which includes a hardware prototype based on a Raspberry Pi system and a software prototype realized in a mobile app.

\subsection{Hardware Prototype} 
We implemented IoT-sensing using Raspberry Pi 4B and deployed several Raspberry Pi 4Bs in one parking lot on campus. The Raspberry Pi consists of 4 $\times$ Cortex-A72 @ 1.5GHz as its CPU and Broadcom VideoCore VI @ 500 MHz as its GPU. The model also includes a b/g/n/ac dual-band 2.4/5 GHz WiFi module, enabling real-time data transmission. We used Ubuntu 20.04 (64-bit) as the Raspbian operating system due to its efficiency compared with the Raspberry Pi OS (32-bit). 

\subsection{Software Prototype}\label{sec:software prototype}
We developed a software prototype for crowdsensing at various parking spaces on the campus. The frontend was implemented using OpenStreetMap with parking spaces marked. RabbitMQ was used to provide connectivity to the backend. The backend was built upon Python Django Rest API, and the cryptographic protocols were implemented using Java Cryptography Library with Py4J Library as a bridge to Django. The CNN model was developed based on Python 3.7, OpenCV 4.2, and TensorFlow 2.3. The training process was done using Google Colab. The detailed specifications of the CNN model are tabulated in Table \ref{tab:CNN_model}. \par

\begin{table}[ht]
  \caption{Summary of the CNN model.}
  \label{tab:CNN_model}
  \begin{tabular}{cc}
     \hline \hline 
    Layer (type)      & Output Shape\\
    \hline
    Input layer     & (64,64,3)\\
    2D convolution layer & (62, 62, 64)\\
    Max pooling 2D layer & (31,31,64)\\
    2D convolution layer & (29, 29, 64)\\
    Max pooling 2D layer & (14,14,64)\\
    2D convolution layer & (12, 12, 64)\\
    Max pooling 2D layer & (6,6,64)\\
    Flatten & 2304 \\
    Dense layer & 128 \\
    Dropout layer & 128 \\
    Dense layer & 1 \\
    \hline \hline 
\end{tabular}
\end{table}

\begin{figure*}[ht]
\centering
\begin{minipage}{\textwidth}
    \centering
    \subfigure[Crowdsourced data]{\includegraphics[width=0.2\textwidth]{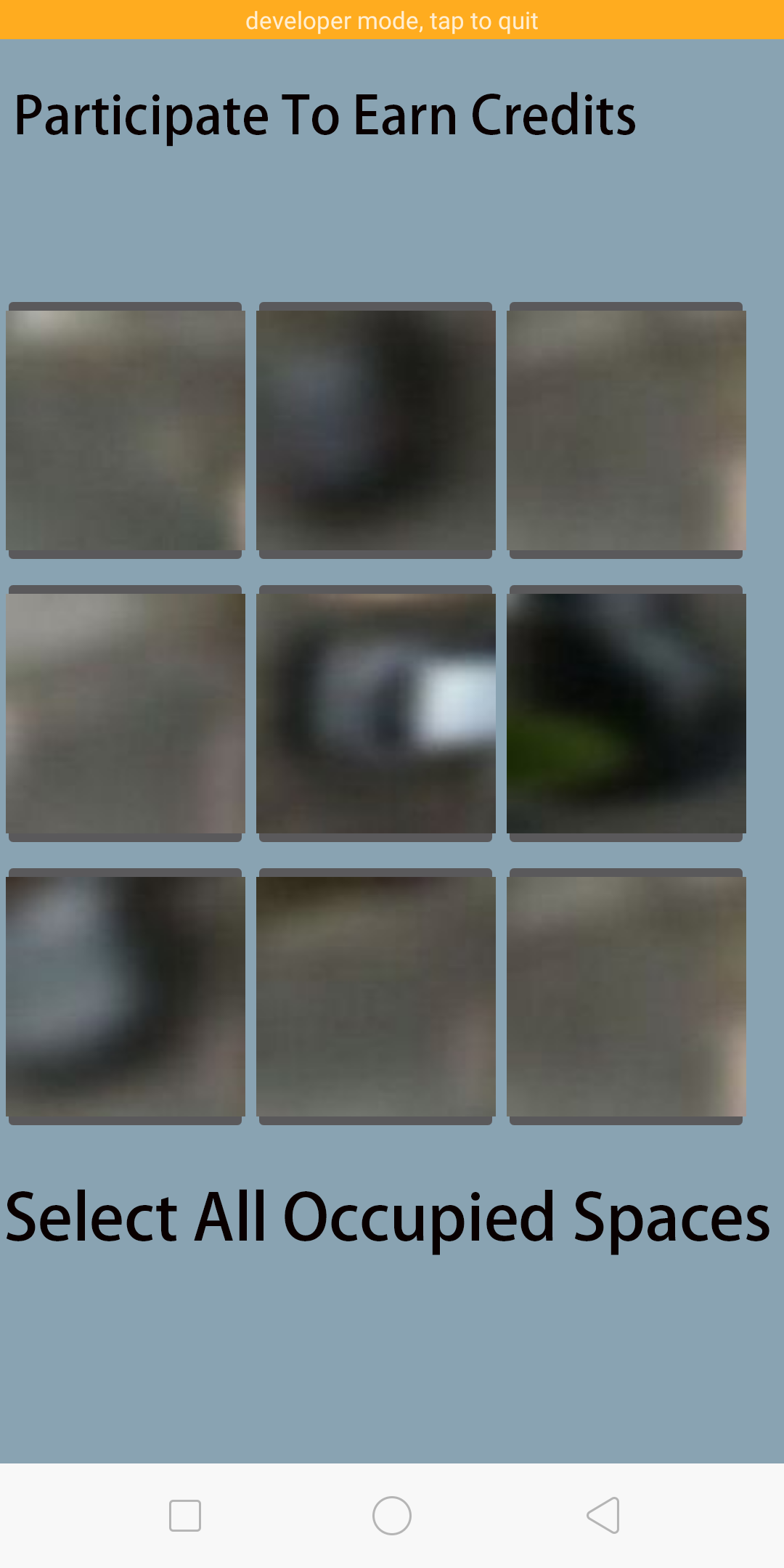}}    
    \subfigure[Crowdsensed data]{\includegraphics[width=0.2\textwidth]{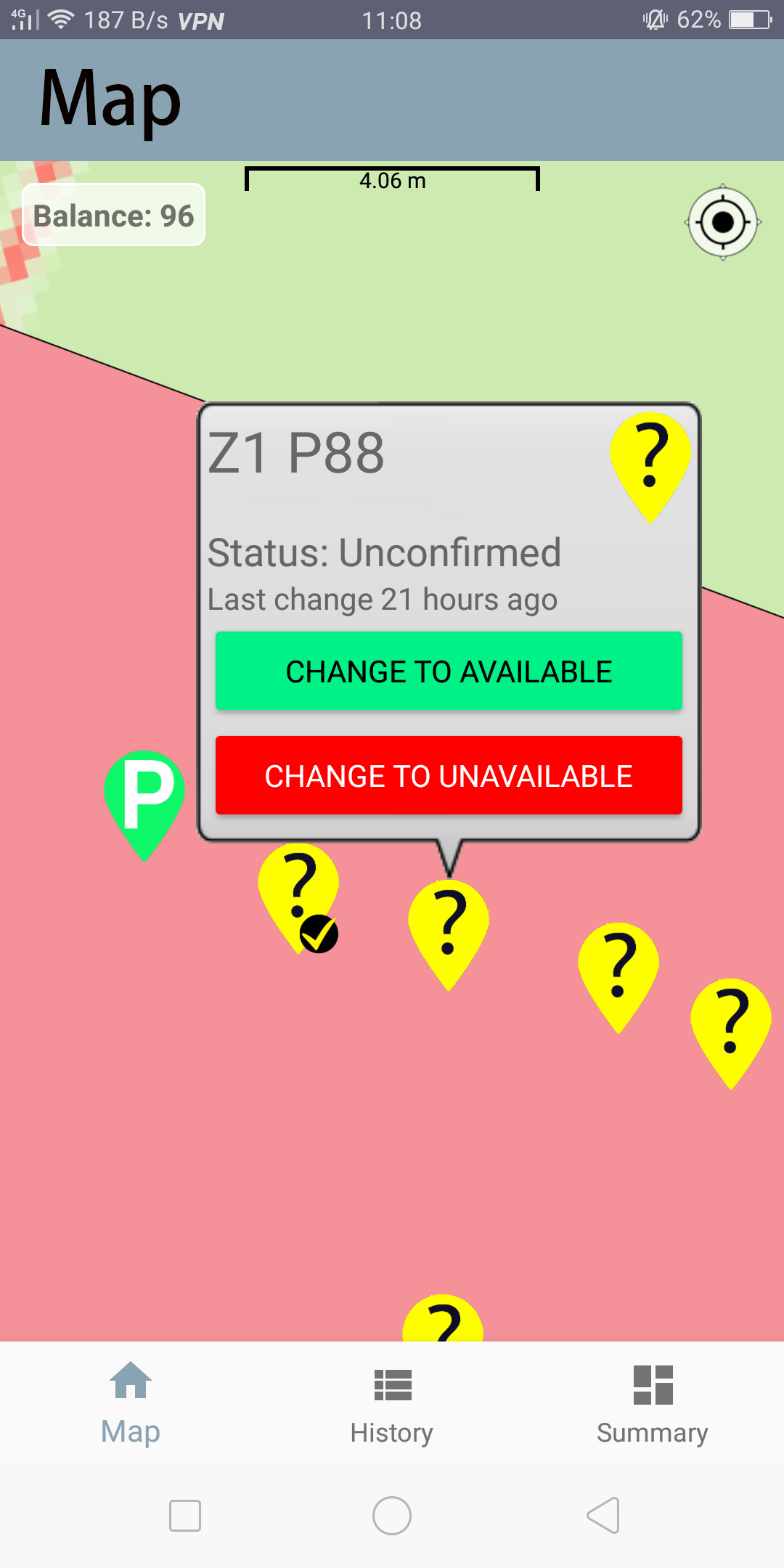}}
    \subfigure[History]{\includegraphics[width=0.2\textwidth]{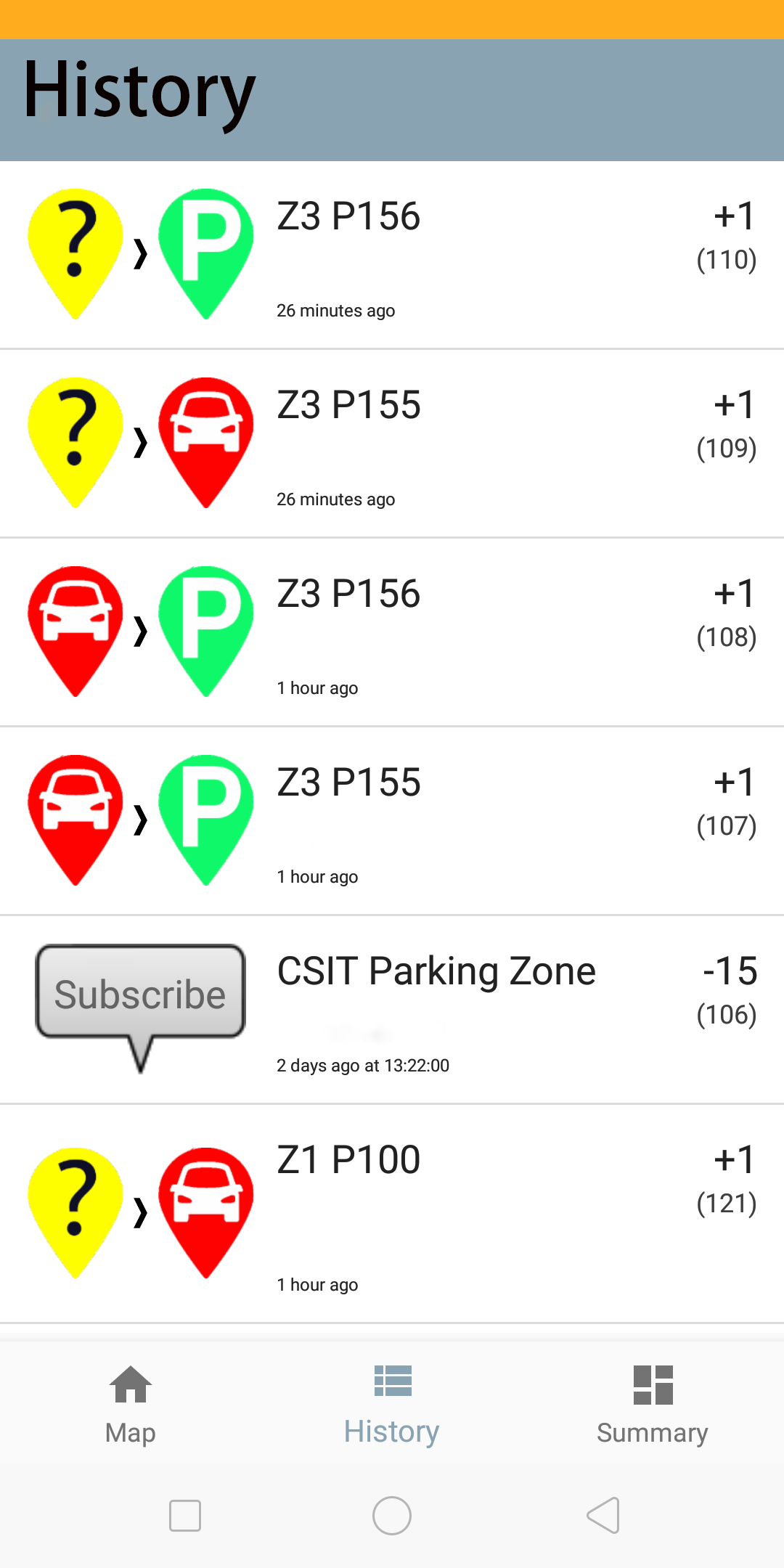}} 
    \subfigure[Credits]{\includegraphics[width=0.2\textwidth]{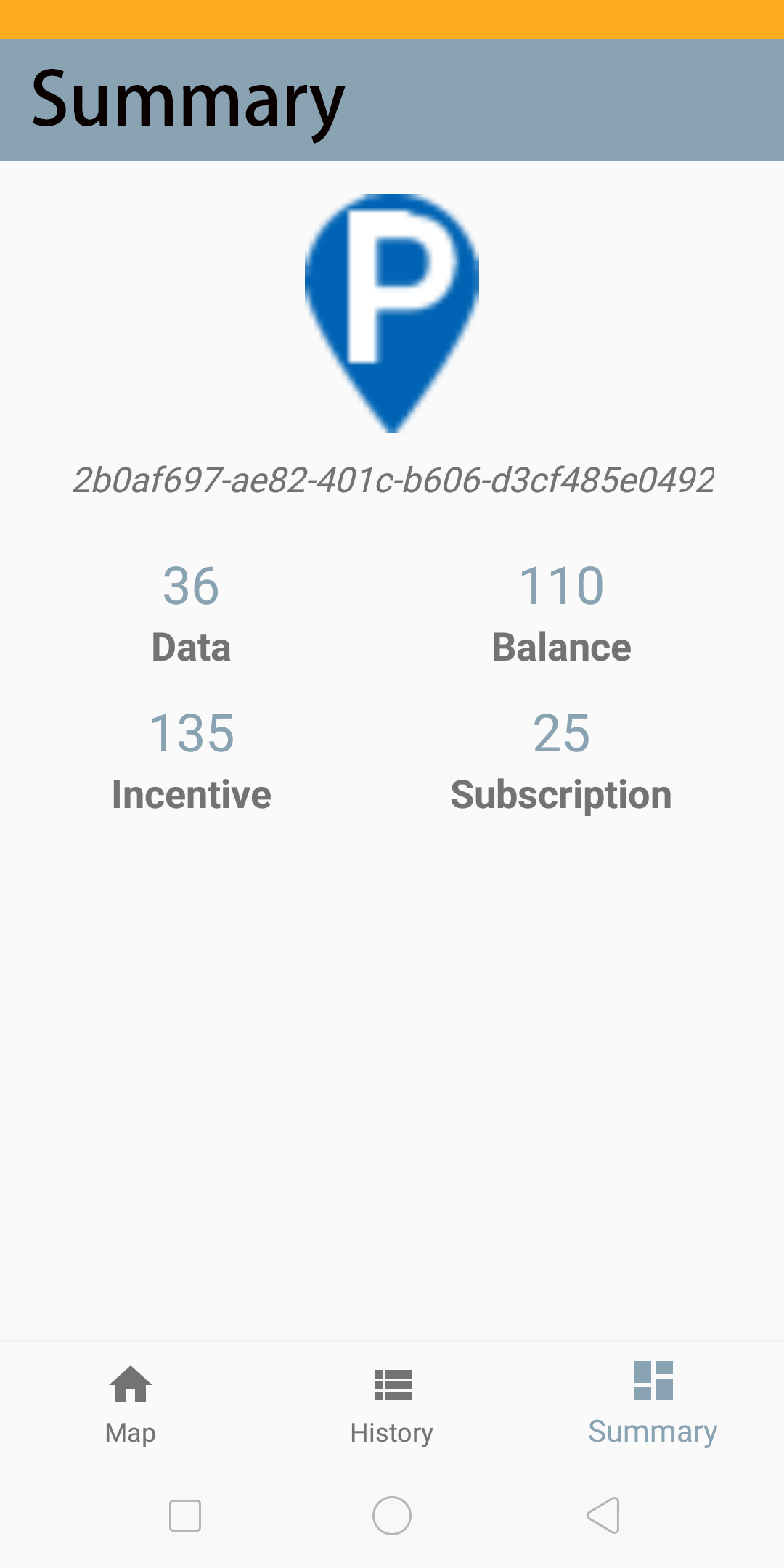}}
 \end{minipage}
     \caption{User interfaces of the mobile app. (a) Submitting crowdsourced training data. (b) Viewing and updating parking space availabilities. (c) Viewing user history.(d) Displaying credit information in an user account.}
    \label{fig:Mobile App}
\end{figure*}

Fig.~\ref{fig:Mobile App} presents the user interfaces of the mobile app. Users are able to view the vacancy information of a particular parking lot on the campus if they are around by using credits. There are two ways to gain credits in the system. Firstly, credits can be gained through crowdsourcing. For instance, a user can contribute to the training set when he identifies the occupied parking spaces within some random selected captured images from the server, as depicted in Fig.~\ref{fig:Mobile App} ($a$). Another way to gain credits is through on-site crowdsensing. For instance, a user can submit the availability information of specific parking spaces, as shown in Fig.~\ref{fig:Mobile App} ($b$). Once the user-submitted data are processed by the server, the parking space will be marked as red if it is unavailable, green if it is available, or yellow if it is unconfirmed \footnote{The status of a parking space will be unconfirmed if the voting results do not reach a majority or there are insufficient votes for a long time.}. Note that we employ a simple majority voting mechanism to determine the status of a parking space. To incentivize participation, users will be credited when the submitted crowdsourced labels are consistent with the majority vote of all submitted crowdsensed data. Moreover, Fig.~\ref{fig:Mobile App} ($c$) depicts the history of user activities, such as reporting availability information and credits used or spent. Finally, Fig.~\ref{fig:Mobile App} ($d$) shows the summary page where users can view their current credits. 

\section{Evaluation Study}\label{sec:evaluation}
This section presents an evaluation study of the feasibility, practicality, effectiveness, and scalability of our hybrid sensing system. We designed and conducted experiments for each component of our system, namely, the IoT-sensing, the crowdsourcing, and the hybrid sensing mobile app. This section is divided into four subsections. In each of the first three subsections, we present an evaluation of a component accordingly, and the last subsection includes a case study of our hybrid sensing system. 

\subsection{Privacy-preserving IoT-Sensing Evaluation} \label{ssec:IoT-sensing eva}
First, we thoroughly investigate the prediction accuracy of our privacy-preserving IoT sensors embedded with an onboard machine learning model considering the effects of different blurriness levels and crowdsourced training datasets. 

\subsubsection{Experimental Setup} \

To simulate the operations in practical scenarios and to investigate the accuracy of our CNN model, we extracted $7,100$ images from a video stream captured by a Raspberry Pi for each blurriness level, as shown in Fig.~\ref{fig:effects}. Among all the images collected, half of them ($3,550$) consisted of cars, which were labeled as cars, and the other half consisted of other objects (e.g., roads, pedestrians, and trees), which were labeled as non-cars. An example of the images collected is depicted in Figure \ref{fig:training_data}. In reality, the on-site images can serve as the crowdsourced data collected from users in our mobile app. In addition, we utilized images from the Stanford Cars dataset \cite{stanforddataset} and the PKLot dataset \cite{pklotdataset} to generate the crowdsourced data for benchmarking the performance of the crowdsourced data collected on-site. Specifically, the Stanford dataset consists of $16,185$ high-quality images of $196$ classes of cars, while the PKLot dataset is made up of $12,416$ extracted images of parking lots. We created a dataset with randomly picked $3,550$ images of cars and $3,550$ images of non-cars from the PKLot dataset. Because the Stanford dataset only contains images of cars, we created another dataset with $3,550$ images of cars from Stanford and $3,550$ images of non-cars from PKLot. For simplicity, we named the datasets as PKLot and Stanford, respectively. It is worth noting that one significant difference between the images from PKLot and Stanford and our on-site images is that images from PKLot and Stanford are not obscured by physical filters. Although the images from PKLot and Stanford may not capture the car features in a specific parking lot, the abundance of these images can be used to benchmark the prediction accuracy of our system as compared to on-site training images, and can bootstrap the training process without a need of sufficient IoT sensors. 

For the on-site training images with the same blurriness level, we divided them into $6$ training sets with different numbers of images, a validation set, and a test set. Particularly, we created training sets with $500$, $1,000$, $2,000$, $3,000$, $4,000$, and $5,000$ images, a validation set with $1,400$ images, and a test set with $700$ images. Note that the training sets with more images contain the training sets with fewer images. For example, the training set with $1,000$ images comprises the training set with $500$ images. For the sake of simplicity, we used the same validation set and test set for each training set under the same blurriness level. For PKLot and Stanford, we repeated the same process when allocating the images, except that the test sets were from our on-site images. As a result, the training image sizes are identical in all datasets. In our hybrid sensing model, we further assume that each user only contributes one image label to the training set. In other words, the number of training images is equivalent to the number of participants submitting crowdsourced data. Finally, we plotted the prediction accuracy for each dataset. 

\subsubsection{Experimental Results and Discussions} \

\begin{figure}[ht]
\centering
\begin{minipage}{0.98\textwidth}
    \centering
    \subfigure[Effect of blurriness and different datasets]{\includegraphics[width=0.48\textwidth]{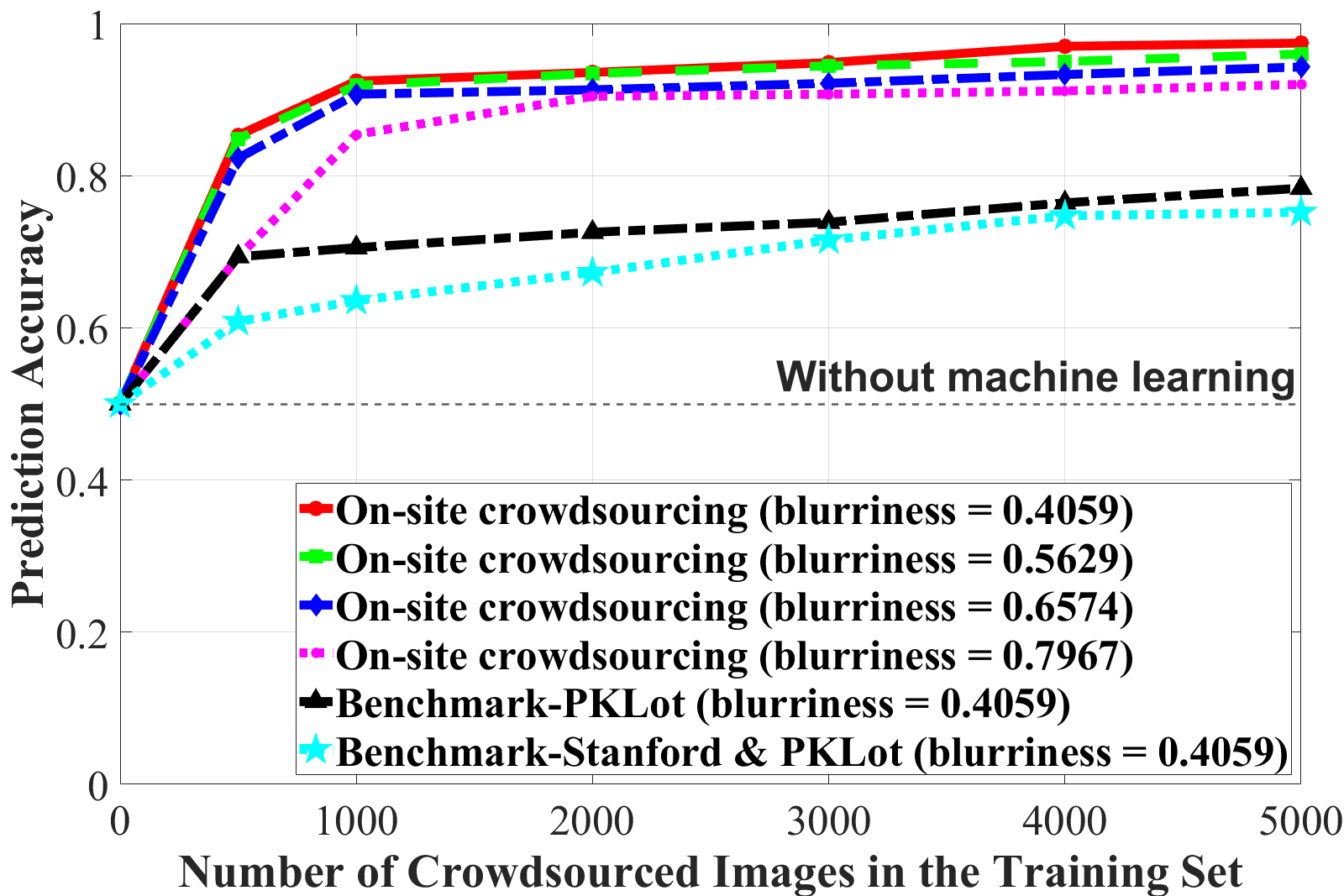}}
    \subfigure[Effect of errors]{\includegraphics[width=0.48\textwidth]{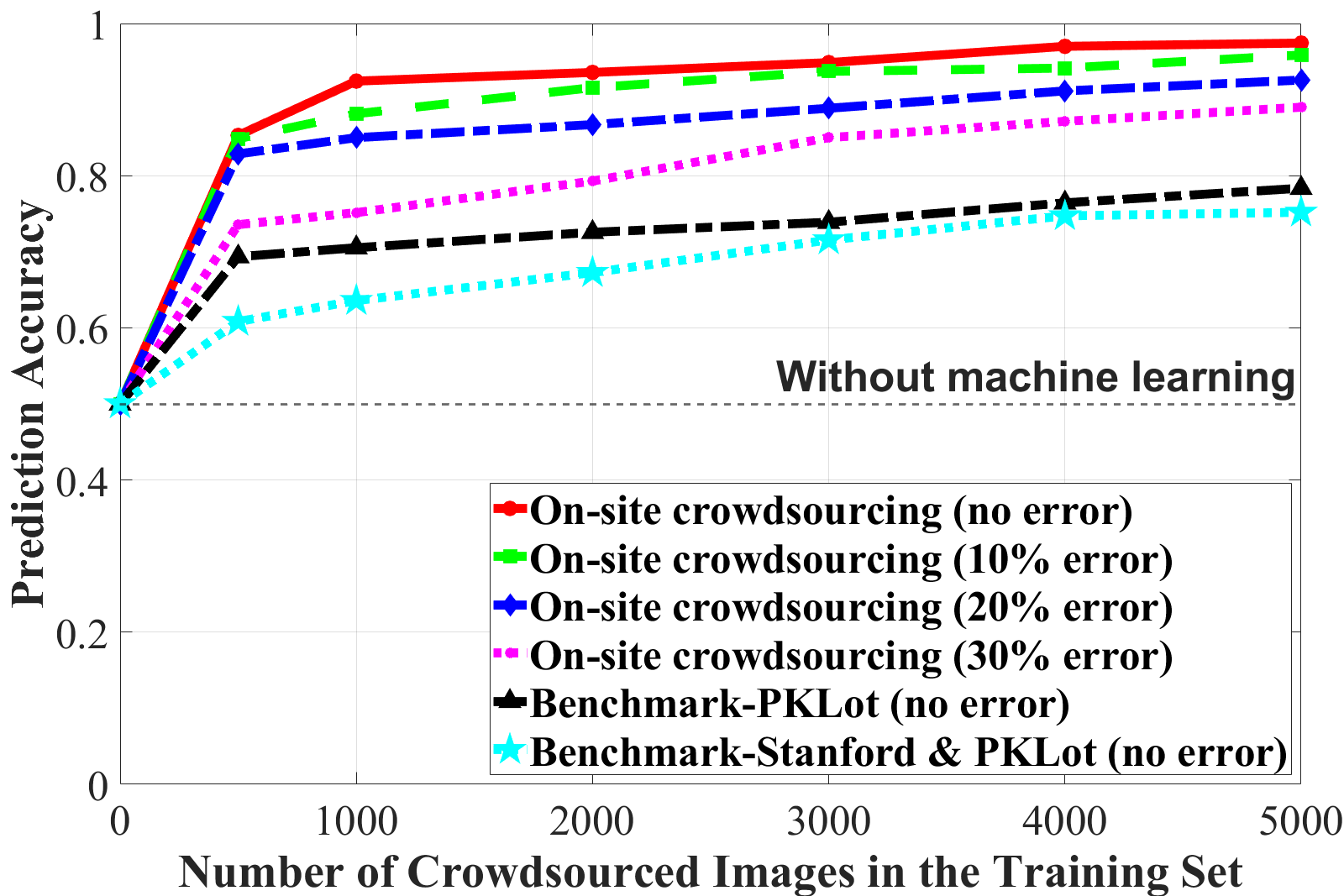}}
 \end{minipage}
     \caption{Prediction accuracy of the CNN model when there are different numbers of crowdsourced data in the training set considering. (a) Different blurriness levels and crowdsourced datasets. (b) Different error rates in the on-site dataset with the blurriness level of 0.4059.}
    \label{fig:accuracy results}
\end{figure}

Fig.~\ref{fig:accuracy results} (a) illustrates the prediction accuracy when there are a different number of crowdsourced images in the training set with different blurriness levels and different datasets. Additional results on training accuracy can be found in the Appendix. For the on-site crowdsourcing datasets with different blurriness levels, it is evident that the prediction accuracy all increases from $0.5$ to above $0.9$ with the initial $2,000$ images. Then, the accuracy converges when there are more crowdsourced data. Moreover, the higher the blurriness is, the lower the prediction accuracy, which aligns with intuition. It is worth mentioning that the CNN model could still achieve a prediction accuracy of $0.92$ even when the blurriness level reaches $0.7967$. This is because although the testing images are highly blurred, the training images have the same blurriness levels, which ensures that the model learns all the blurred features. In contrast, if the testing images are highly blurred but the training images are sharp, the prediction accuracy will decrease more (as illustrated in the Appendix). As a result, both online datasets perform worse than the on-site ones even though the blurriness level in the test images is only $0.4059$. However, the model using both datasets can still reach an accuracy of almost $0.8$, which we believe is sufficient for most real-world hybrid sensing models to start with. Therefore, we remark that obtaining online crowdsourced training images is applicable and practical when the IoT sensors need to be deployed on a large scale within a short time. In practice, to further improve the prediction accuracy of a specific sensor in a parking lot, we can collect more images from that parking lot. In our hybrid sensing system, more on-site images can be obtained through user contribution, as demonstrated in Fig.~\ref{fig:Mobile App} (a).

\subsection{Crowdsourcing Benefits to IoT-Sensing}
In our hybrid sensing system, there is a need to carefully understand the potential benefits of crowdsourcing to enhance the accuracy of IoT-sensing. The accuracy of the IoT sensors depends directly on the quality of the training data. We have identified three possible scenarios where mislabeled data may exist in the training datasets: (1) malicious users may submit wrong information deliberately, (2) honest users may submit wrong information accidentally, and (3) when the crowdsourced images are highly blurred, it is difficult to differentiate between cars and non-cars. Therefore, it is crucial to understand how inaccurate data may impact our CNN model. 

To simulate the errors obtained from the crowd, we adopted the same on-site training sets, validation sets, and test sets as in Section \ref{ssec:IoT-sensing eva} but randomly interchanged 10\%, 20\%, and 30\% of the labels (i.e., labeling cars as non-cars and vice versa) in the training sets. Similarly, we plotted the prediction accuracy of the CNN model using $500$, $1,000$, $2,000$, $3,000$, $4,000$, and $5,000$ training images.

Fig.~\ref{fig:accuracy results} (b) displays the prediction accuracy of the CNN model when there are different levels of error percentages in the training set with a blurriness level of 0.4059. The prediction accuracy with various error rates under all blurriness levels can be found in the Appendix. It is evident that the prediction accuracy with different error percentages all shows a rapidly growing trend when the number of crowdsourced images available in the training set increases from $0$ to $1,000$, and gradually converges to $90\%$ or above when more training images become available. Also, it is noticeable that more errors there are in the crowdsourced images, the lower the prediction accuracy. However, when there are $30\%$ errors in the training set, the prediction accuracy can still reach almost $90\%$, which is around $10\%$ higher than the benchmarking datasets with no mislabeling error. This implies that our CNN model is capable of handling noises and errors in the training sets well. Additional experimental results are plotted according to different levels of blurriness in the Appendix. 

\subsection{Performance of Hybrid Sensing System}
Next, we evaluate the performance of our hybrid sensing app. We recorded the running time, the  communication overhead, and the parking availability of our protocol between a server and a user device. We used a Raspberry Pi 4B with Quad-core Cortex-A72 and 4 GB of RAM as the user device. A Linux laptop with a 1.8 GHz 4-Core Intel Core i7-8550U processor, and 8 GB of RAM was used as the server. All the results were averaged over 10 instances.

\subsubsection{System Running Time}\

We measured the running time of each stage in our protocol. Note that the number only reflects the time taken for computation without considering network latency. As shown in Fig.~\ref{fig:running time and overhead} (a), all the data in each stage could be processed within a short time. Specifically, the running time in the registration stage is the longest among all stages, with a value of around $0.4$ seconds in total. This is because the stage involves generating $2$ secret keys $s$ and $s'$ and computing $4$ commitments ${\tt Cm}(s'), {\tt Cm}(q), {\tt Cm}(s)$, and ${\tt Cm}(b_0)$. The running time of the rest stages is relatively small (all under $0.2$ seconds), with the data submission stage being the shortest among all stages, which costs a total of $0.1$ seconds. Moreover, it is noticeable that the running time on the server-side is slightly shorter than the user-side for most stages since the computing power of the Raspberry Pi is lower than the laptop, and users need to generate more information and send it to the server, as illustrated through the communication overhead shown below. 

\subsubsection{Communication Overhead} \

\begin{figure}[ht]
\centering
\begin{minipage}{0.98\textwidth}
    \centering
    \subfigure[Running time]{\includegraphics[width=0.48\textwidth]{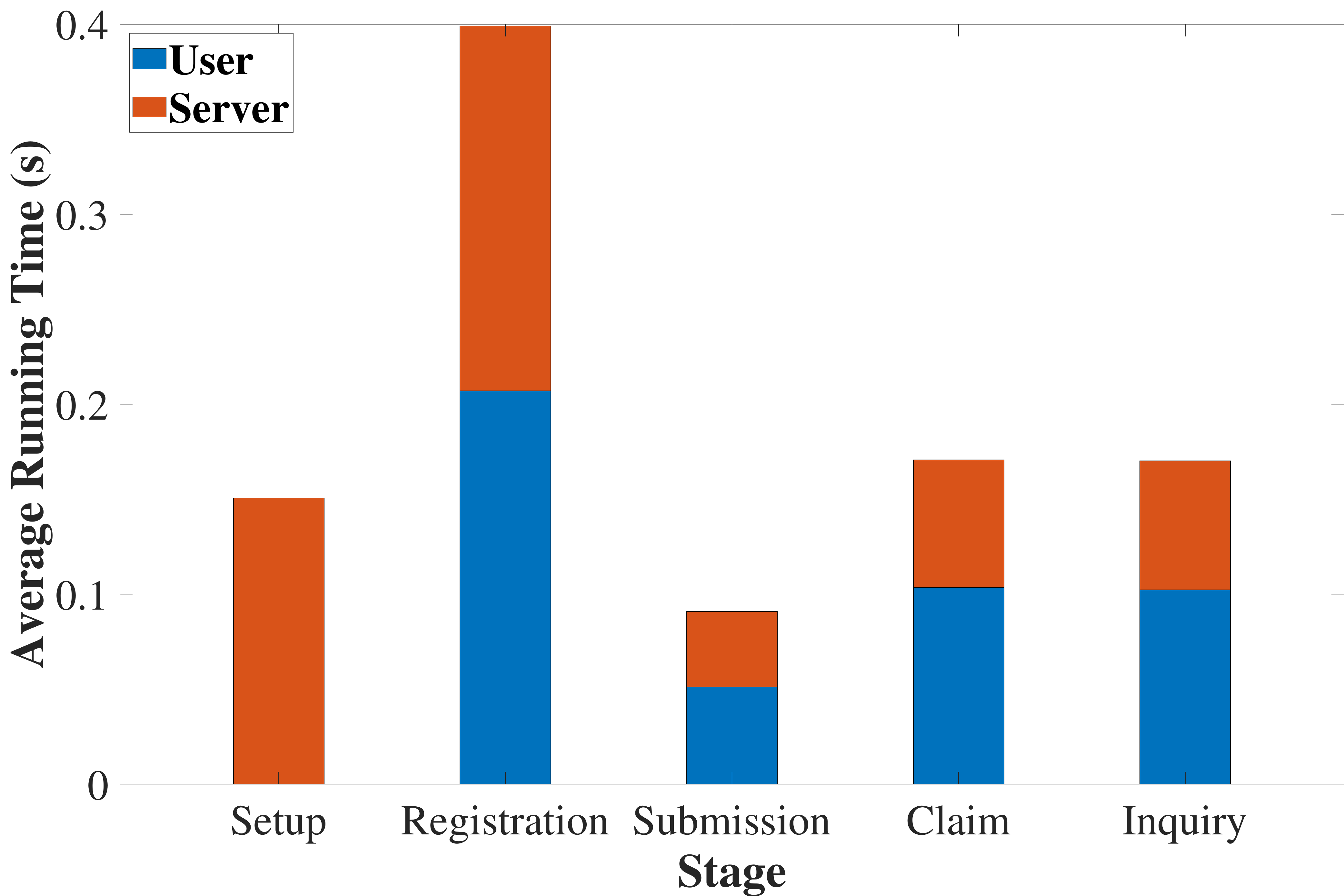}}
    \subfigure[Communication overhead]{\includegraphics[width=0.48\textwidth]{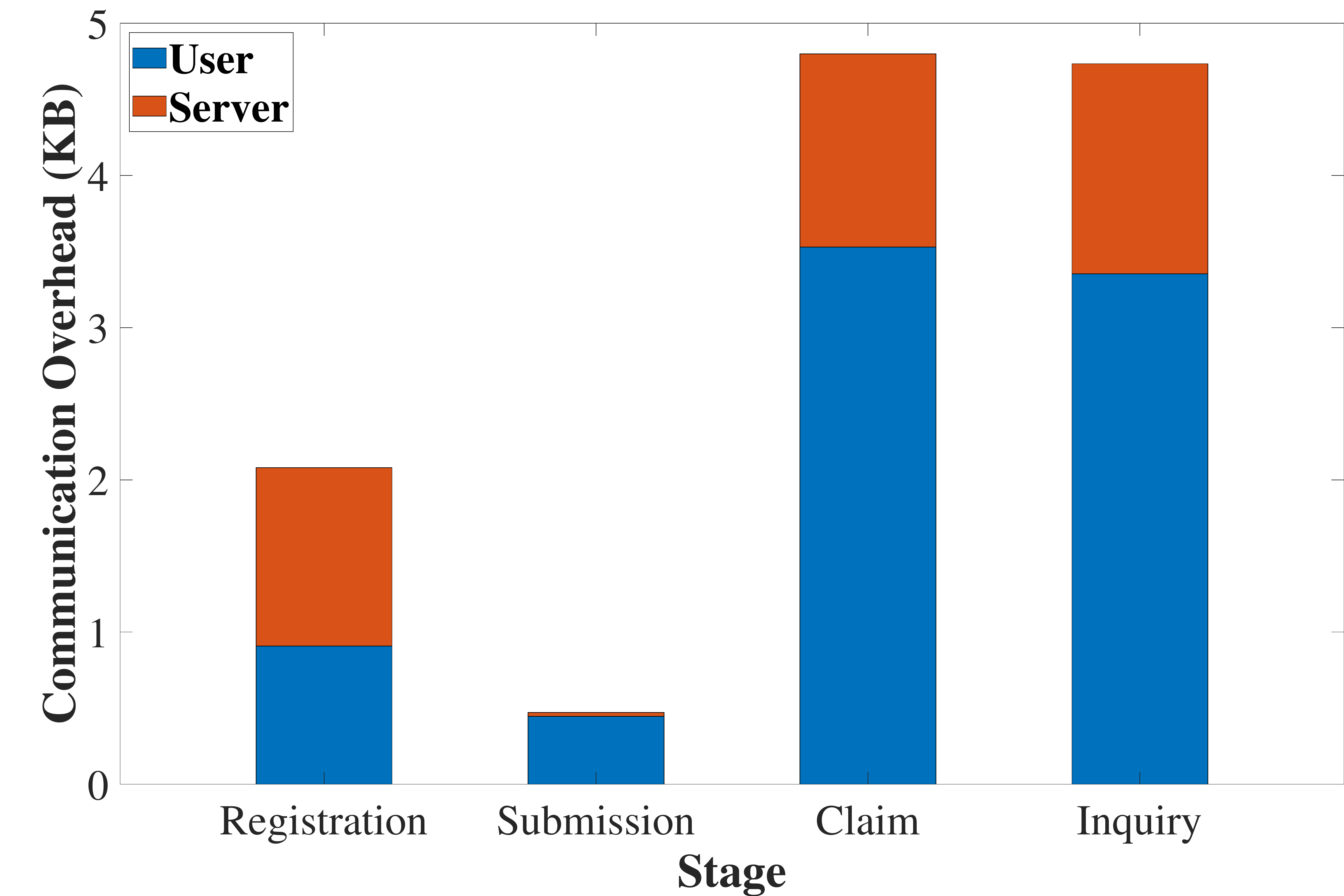}}
 \end{minipage}
     \caption{Communication and computation cost of the hybrid sensing app. (a) The running time in different stages in the protocol. (b) The communication overhead in different stages in the protocol.}
    \label{fig:running time and overhead}
\end{figure}

Fig.~\ref{fig:running time and overhead} (b) shows the average total volume of the transmission data in the registration, data submission, credit claiming, and inquiry stages. Note that the setup stage does not incur any communication cost. We observe that the overhead in all stages are quite small, and the overhead generated from the server-side are much smaller than the user-side in all stages except registration ($0.9$ KB by the user and $1.2$ KB by the server) because users need to generate and send non-interactive zero-knowledge proofs. In addition, it is evident that the data submission stage incurs the least total overhead, with a size of about $0.5$ KB, while the credit claiming stage and inquiry stage generate the most data, with a size of around $4.8$ KB and $4.7$ KB, respectively. Hence, the total data volume in the credit claiming and inquiry stage dominates the entire protocol. 

\subsubsection{Confirmation Time of Crowdsensing} \

\begin{figure}[ht]
    \centering
	\includegraphics[width=0.5\textwidth]{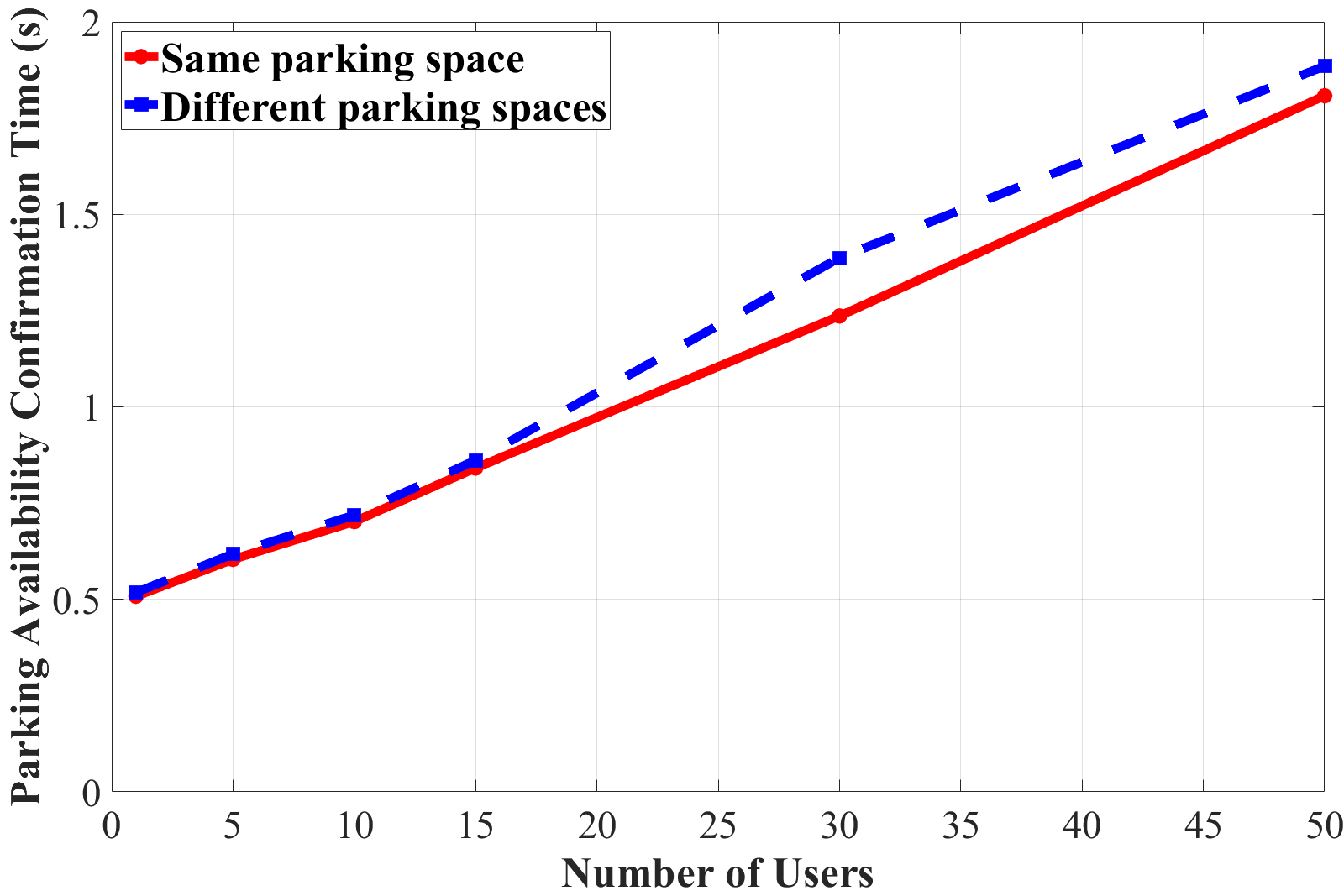}
	\caption{Parking availability confirmation time when different number of users submit crowdsensed information simultaneously.}
	\label{fig:confirmation time}
\end{figure}

To evaluate the performance of our privacy-preserving crowdsensing system, we focus on one crucial performance metric - the confirmation time of crowdsensed data in the server considering both IoT-sensing and crowdsensing. To simulate data arrivals in IoT-sensing and crowdsensing, we randomly generated $1, 5, 10, 15, 30$, and $50$ users and let those users send random availability information to the server simultaneously for the same parking space or a number of different parking spaces. Note that the IoT sensors, which regularly publish availability information, can serve as users and contribute crowdsensed data as well. Fig.~\ref{fig:confirmation time} displays the total parking availability confirmation time incurred. The confirmation time demonstrates a linearly growing trend with increased users. When all users submit parking availability information for one parking space, the confirmation time starts from around $0.51$ seconds with $1$ user to around $1.81$ seconds with $50$ users. On the other hand, when users submit availability data for different parking spaces, the total confirmation time will be slightly higher, ranging from $0.52$ ms ($2\%$ more) to $1.88$ ms (around $4\%$ more). Overall, we observe that our hybrid sensing app can process users' data and update the availability information accordingly within a short period, even when  several users are sending requests at the same time.

\subsection{A Hybrid Sensing Case Study}
Finally, we present a case study of our hybrid sensing system to study the benefits of IoT-sensing and crowdsensing. We considered a parking lot with $4$ parking spaces (namely, parking spaces $1$ - $4$) where parking spaces $1$ and $2$ could be covered by both IoT-sensing and crowdsensing. In contrast, parking spaces $3$ and $4$ could only be covered through crowdsensing. The setting was designed on purpose to simulate the scenario where some parking spaces in a parking lot cannot be detected by cameras due to coverage issues. We deployed a Raspberry Pi-based camera obscured by a blurry filter with a blurriness level of $0.4059$. Meanwhile, we asked several volunteers to observe and contribute to the crowdsensed data for all parking spaces. We plotted the parking availability over time for each parking space and the aggregated availability for the whole parking lot. The overall parking availability was aggregated based on whether at least one parking space was available at a given time. The study was conducted during $9$ am and $5$ pm on a workday on our campus. 

\begin{figure}[ht]
    \centering
	\includegraphics[width=0.98\textwidth]{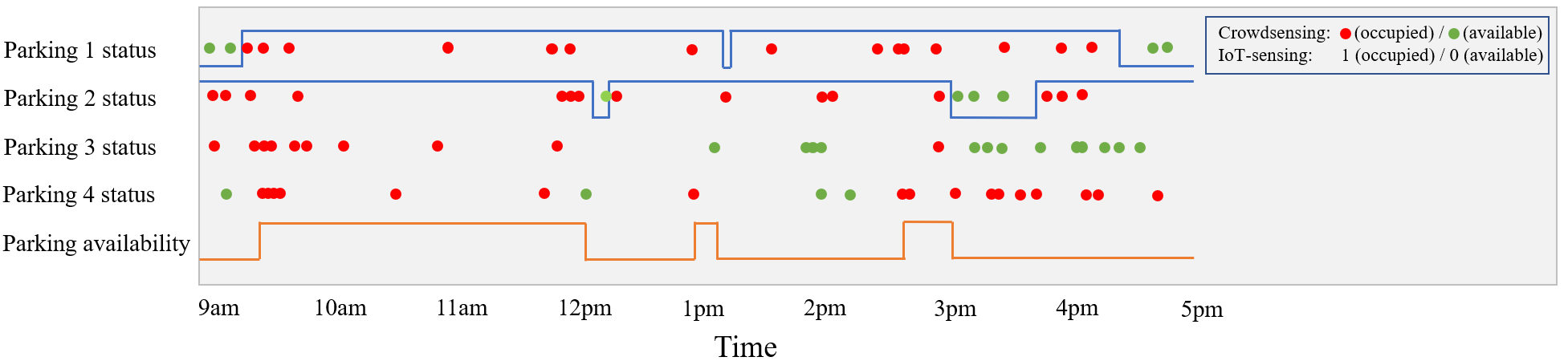}
	\caption{A case study showing the parking availability information of a parking lot from 9am to 5pm.}
	\label{fig:case study}
\end{figure}

As shown in Fig.~\ref{fig:case study}, in parking spaces $1$ and $2$, the IoT sensor could provide continuous sensing data as illustrated in the binary sequences where a status of $1$ represents occupied, and a status of $0$ represents available. In contrast, the availability of parking spaces $3$ and $4$ was only reported by intermittent crowdsensing data, represented by red and green points. Due to the nature of data arrivals (i.e., IoT-sensing provides continuous data and crowdsensing provides intermittent data), it is obvious that IoT sensors can provide better reliability when the availability of a parking space changes but there is no crowdsensed data. For example, when parking space $2$ became available right after $3$ pm, the IoT sensor could detect it immediately, whereas there was a short delay before the arrival of the next crowdsensed data. Consequently, the overall parking availability could capture the sudden change shortly. However, although crowdsensed data are less reliable compared to IoT-sensed data, crowdsensing can provide a high sensing coverage. For instance, parking spaces $3$ and $4$ could only be covered through crowdsensing in our case study, and in practice, crowdsensing can reach more parking lots where IoT sensors are not yet available. 

\section{Conclusion}\label{sec:conclusion}
This paper presents a privacy-preserving hybrid sensing system with an application for smart parking availability monitoring. We integrated IoT-sensing with crowdsensing, enhanced with privacy-preserving techniques. In particular, we obscured IoT sensors with physical blurry filters in IoT-sensing and applied a cryptographic solution based on cryptographic commitments, zero-knowledge proofs, and anonymous credentials in crowdsensing. We also trained a machine learning model for parking recognition under blurry filters using crowdsourcing. Proof-of-concept prototypes, including a Raspberry Pi system and a mobile app, were developed, and an evaluation study of the machine learning model and the effects of crowdsourcing were presented in the article. 

In future work, we will address further research challenges. For instance, how to extend the functionalities of a hybrid sensing system to the setting of heterogeneous sensors. In particular, we will consider using multiple cameras with overlapping fields of view to improve the accuracy of machine learning object recognition. Furthermore, we will study how to apply more recent deep learning techniques to our Raspberry Pi camera system. Meanwhile, we will investigate the impact on the detection accuracy under different environmental conditions, such as lighting and weather. Additionally, we will study a well-balanced incentive mechanism for earning credits from crowdsensing that will dynamically adjust the credit rates in response to the behavior of users.

\section*{Acknowledgment}
We are grateful to the helpful comments and suggestions offered by the associate editor and anonymous reviewers. This project was supported by ARC Discovery Project No: GA69027/DP200101985.

\bibliographystyle{ACM-Reference-Format}
\bibliography{references}

\appendix
\newpage

\appendix

\section{Additional Experimental Results}\label{sec:complete exp}

\begin{figure}[ht!]
\centering
\begin{minipage}{0.8\textwidth}
    \centering
    \subfigure[]{\includegraphics[width=0.48\textwidth]{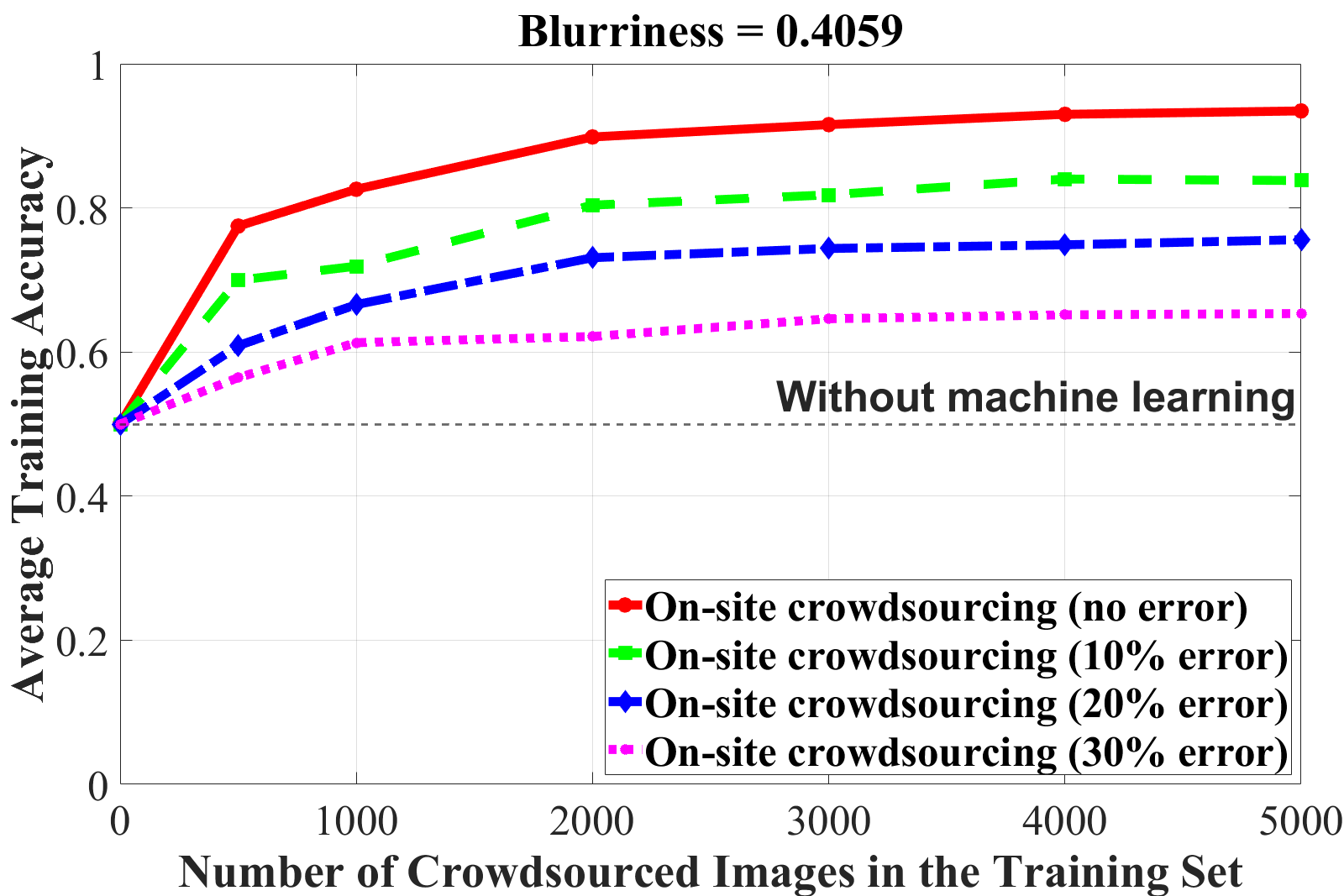}}    
    \subfigure[]{\includegraphics[width=0.48\textwidth]{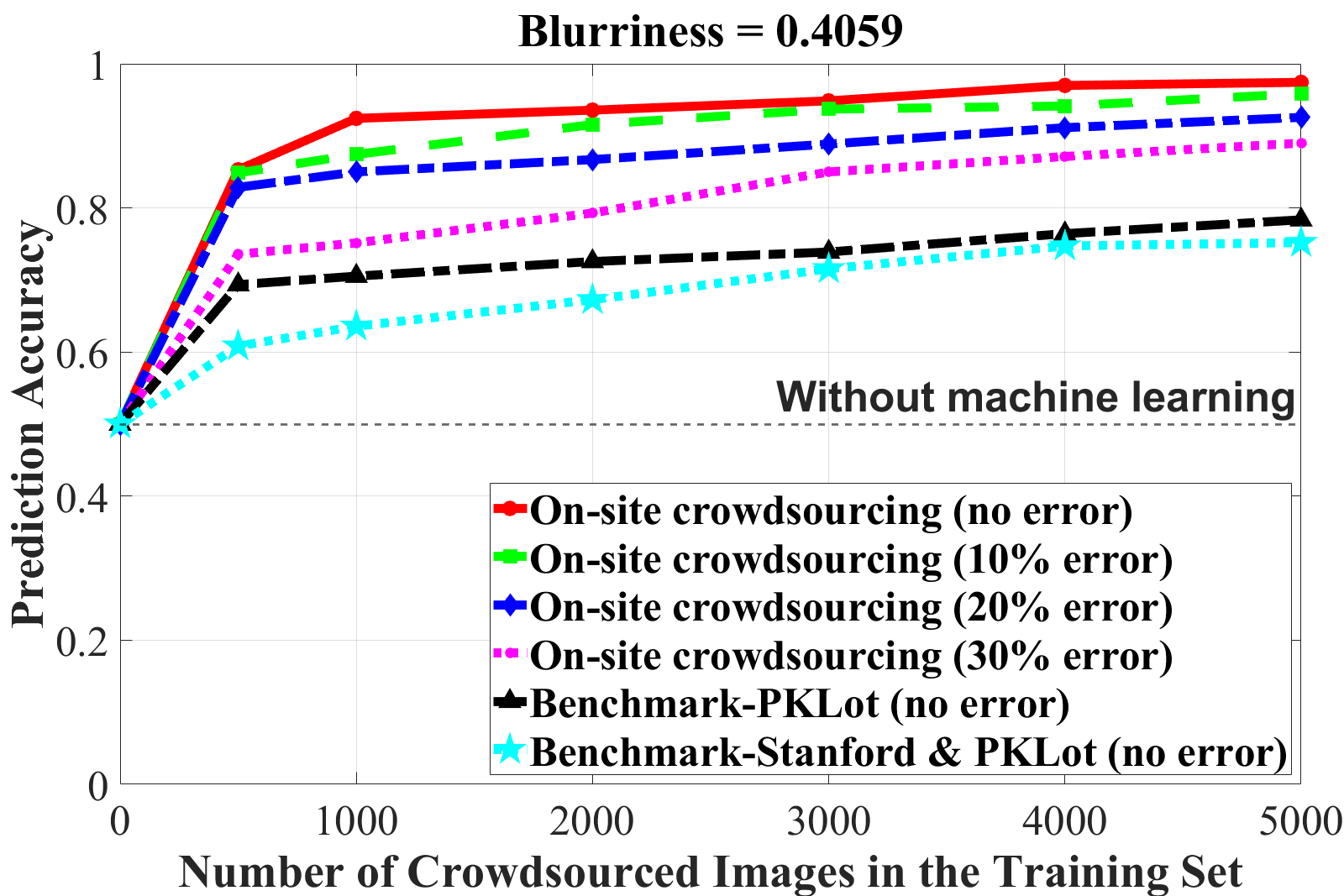}}\qquad
     \medskip
    \subfigure[]{\includegraphics[width=0.48\textwidth]{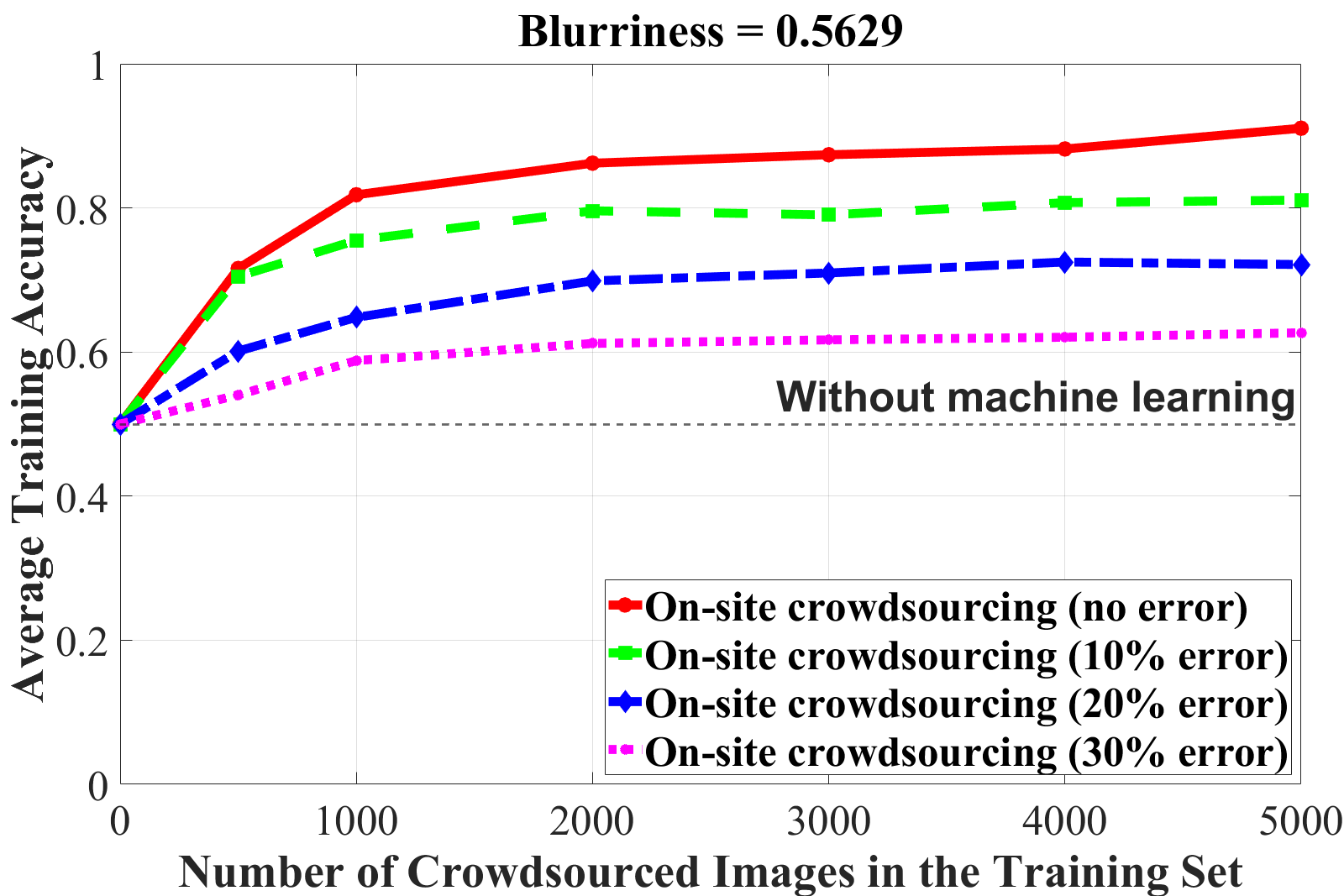}} 
    \subfigure[]{\includegraphics[width=0.48\textwidth]{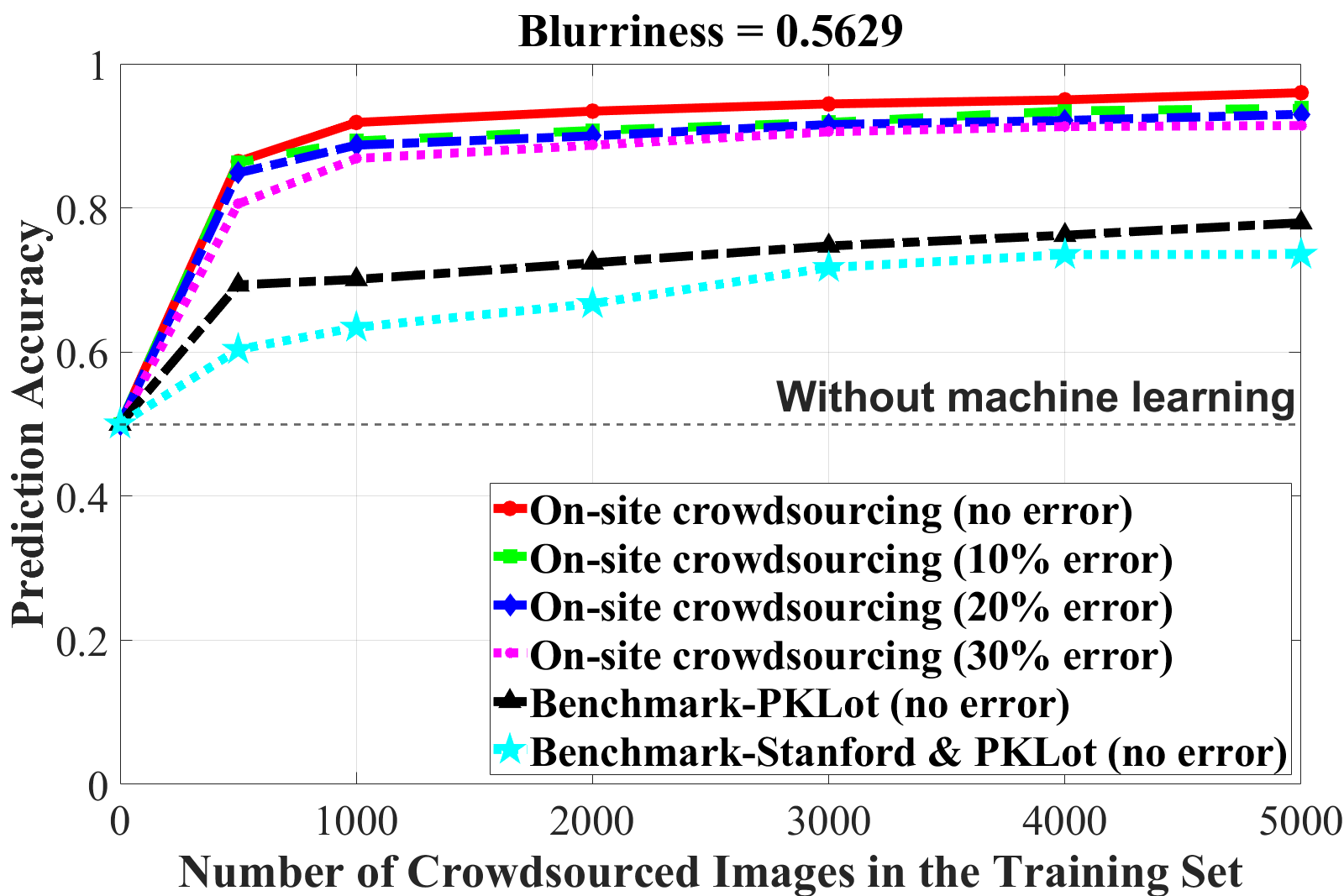}}\qquad
    \medskip
    \subfigure[]{\includegraphics[width=0.48\textwidth]{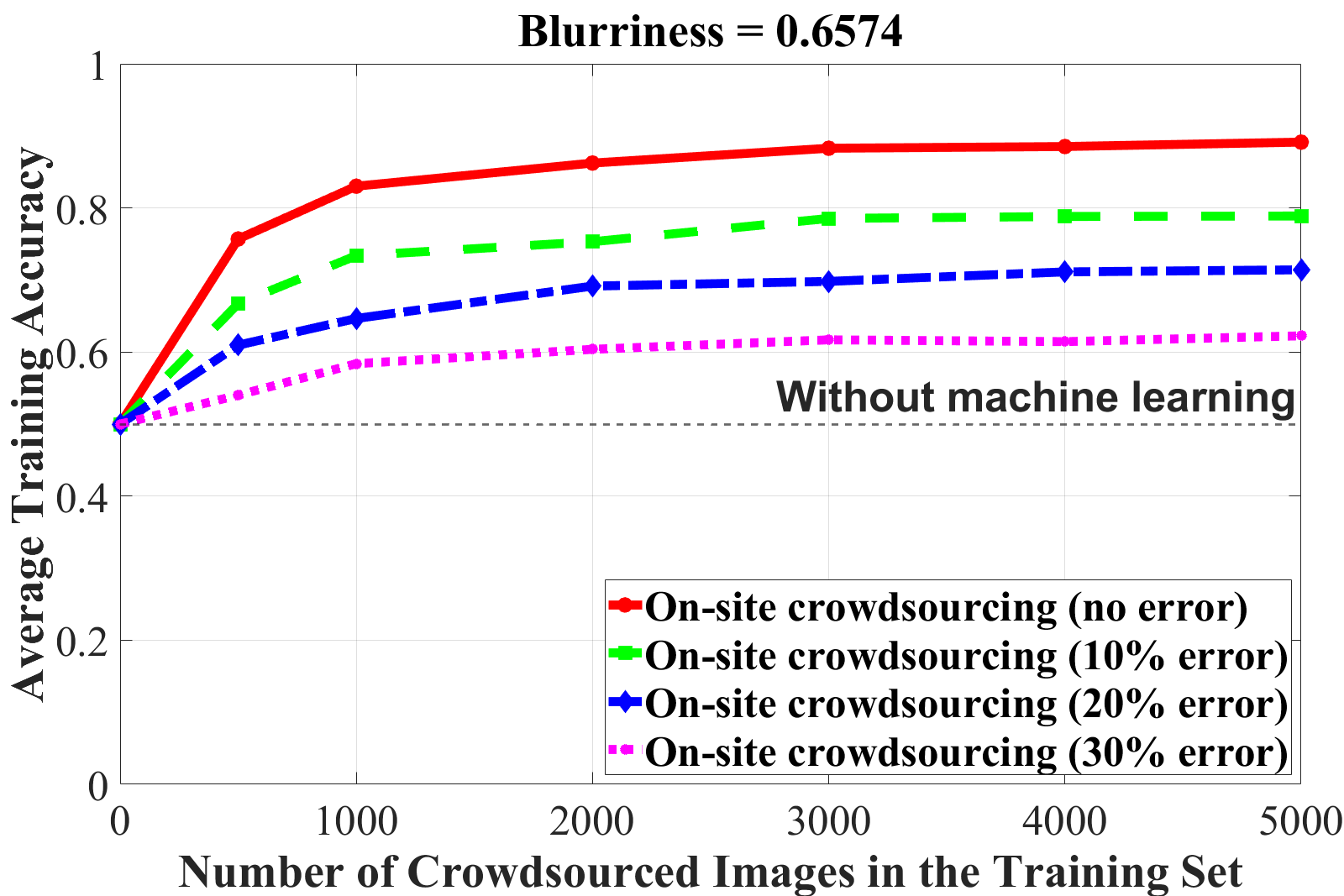}} 
    \subfigure[]{\includegraphics[width=0.48\textwidth]{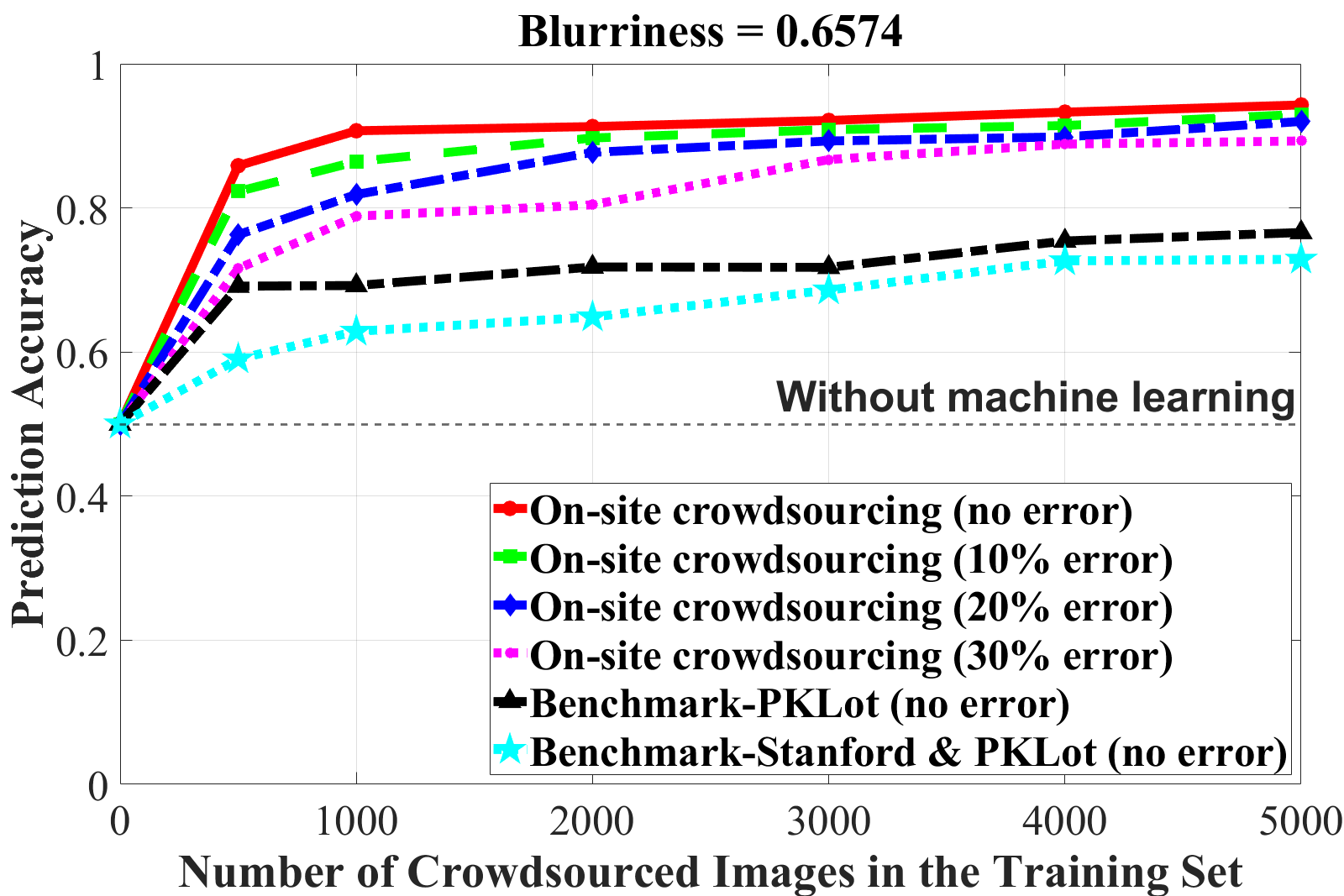}}\qquad
    \medskip
    \subfigure[]{\includegraphics[width=0.48\textwidth]{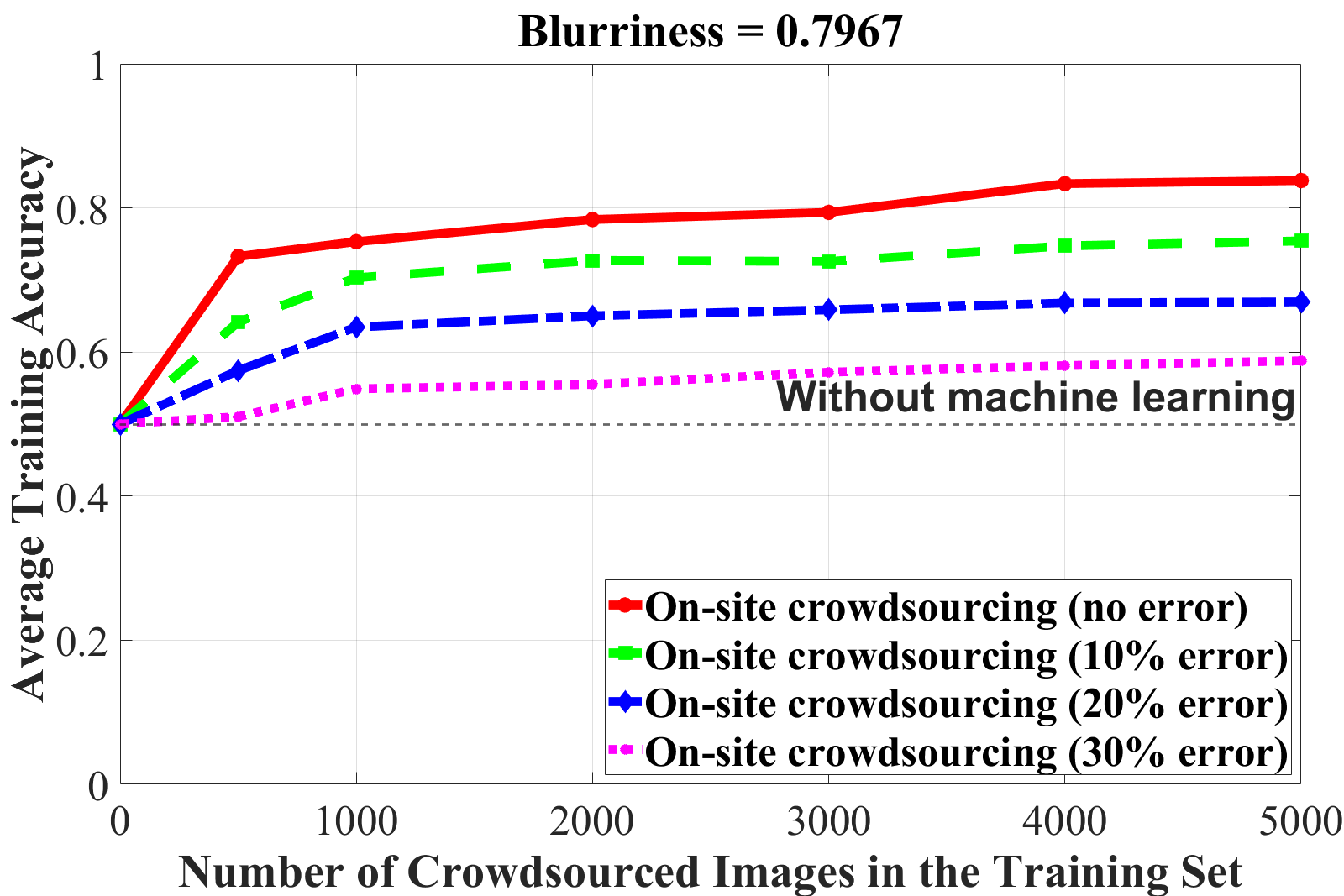}} 
    \subfigure[]{\includegraphics[width=0.48\textwidth]{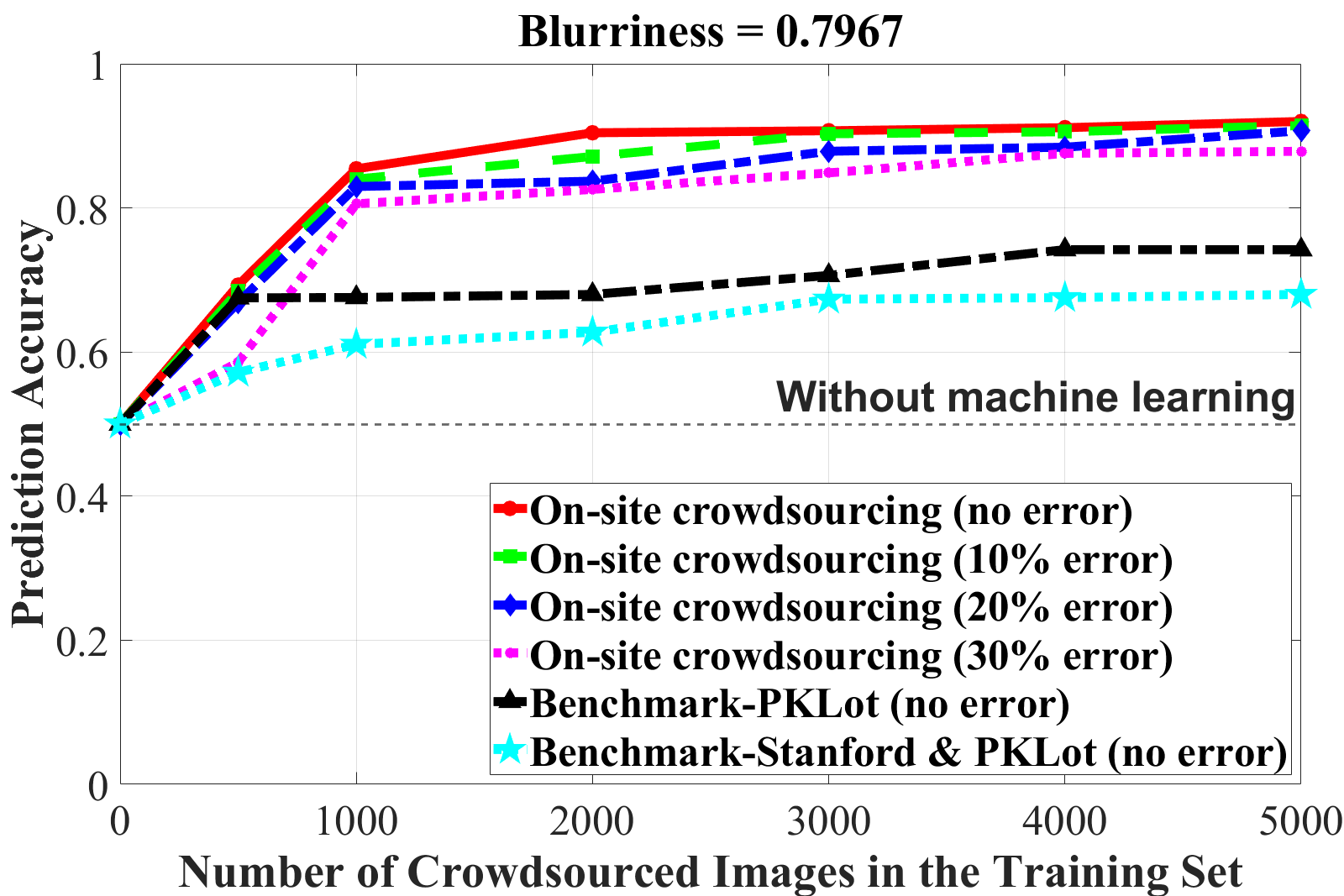}}
 \end{minipage} \vspace{-20pt}
     \caption{Average training accuracy and prediction accuracy vs. number of crowdsourced data at different blurriness levels using different datasets. The average training accuracy is calculated out of 50 epochs.}
    \label{fig:complete results}
\end{figure}

\newpage

\section{Security of Zero-Knowledge Proofs}\label{sec:appendix}

We first prove the security of the $\sigma$-protocol-based zero-knowledge proofs (i.e., {\tt zkpCm}, {\tt zkpMbs} and {\tt zkpNN}) by proving their completeness, soundness and zero-knowledge. 
\subsection{Completeness Proof}
It is straightforward to prove the completeness as one only needs to check if the verifier is convinced if an honest prover correctly executes the protocol.  
\begin{itemize}
    \item ({\tt zkpCm}): We just need to show whether $g^{z_x} \cdot h^{z_r} \overset{?}{=}  Cm(x', r') \cdot Cm(x, r)^\beta$. Because $z_x = x' + \beta \cdot x$ and $z_r = r' + \beta \cdot r$, it is evident that $g^{x'+\beta x} \cdot h^{r'+\beta r} = g^{x'+\beta x} \cdot h^{r'+\beta r}$.
    \item ({\tt zkpMbs}): We need to prove whether $\beta \overset{?}{=} \sum_{i=1}^n \beta_j$ and $g^{z_{x_j}} \cdot h^{z_{r_j}} \overset{?}{=} Cm(x'_j, r'_j) \cdot \Big( \frac{Cm(x, r)}{g^{x_j}} \Big)^{\beta_j}$. It's easy to prove the former equation because $\beta_i = \beta - \sum_{j\ne i}\beta_j$. For the latter one, when $j\in\{1,...,n\}\backslash\{i\}$ we can get $g^{x'_j+(x_i-x_j)\beta_j} \cdot h^{r'_j+r \cdot \beta_j} \overset{?}{=} g^{x'_j} \cdot h^{r'_j} \cdot g^{(x-x_j)\beta_j} \cdot h^{r\cdot \beta_j}$ and hence $g^{(x_i-x_j)\beta_j} \overset{?}{=} g^{(x-x_j)\beta_j}$. The equality is achieved because $x=x_i$. Similarly, when $x=j$ we can get $g^{x'_i-x'_j} \overset{?}{=} g^{(x-x_j)\beta_j}$. The equality is achieved because $x=x_i=x_j$.
    \item ({\tt zkpNN}): We need to prove whether $h^{z_r} \overset{?}{=}  Cm(0,r') \cdot Cm(x, r)^{-\beta}\cdot\prod_{i=1}^m Cm(b_i, r_i)^{\beta\cdot 2^{i-1}}$. Substituting $z_r = r' + \beta \cdot (\sum_{i=1}^m r_i \cdot 2^{i-1} - r)$ into the equation, we can get $h^{ r' + \beta \cdot (\sum_{i=1}^m r_i \cdot 2^{i-1} - r)} \overset{?}{=}  h^{r'} \cdot g^{-x\beta} \cdot h^{-r\beta} \cdot g^{\sum_{i=1}^m b_i\cdot \beta\cdot 2^{i-1}} \cdot h^{\sum_{i=1}^m r_i\cdot \beta\cdot 2^{i-1}}$. Note that the product of the commitments can be changed to a summation due to the homomorphic property. Finally, we can get $1 \overset{?}{=} g^{\beta \cdot(\sum_{i=1}^m b_i \cdot 2^{i-1}-x)}$, which is easy to verify because $\sum_{i=1}^{m}b_i \cdot 2^{i-1} = x$.
 
\end{itemize}

\subsection{Soundness Proof}

To prove soundness, we require a \textit{Knowledge Extractor} that can extract the knowledge the prover who wants to prove by making the prover successfully answer two random challenges $\beta_1$ and $\beta_2$, specifically:

\begin{itemize}
    \item ({\tt zkpCm}): The extractor sends a random challenge $\beta_1$ to the prover, and the prover replies with $z'_x = x'+\beta_1$ and $z'_r=r'+\beta_1$. Then, the extractor rewinds the execution with a new challenge $\beta_2$, and the prover replies with $z''_x = x'+\beta_2$ and $z''_r=r'+\beta_2$. Note that because of rewinding, the prover will always send the same $x'$ and $r'$ during each execution. Finally, the extractor is able to extract the prover's secret  by computing $x=\frac{z'_x-z''_x}{\beta_1-\beta_2}$ and $r=\frac{z'_r-z''_r}{\beta_ta_2}$.
    \item ({\tt zkpMbs}): The extractor sends two random challenges $\beta_1$ and $\beta_2$ before and after rewinding, and the prover replies with $z_{r_j}=r'_j+r \cdot \beta_1$ and $z'_{r_j}=r'_j+r \cdot \beta_2$. Next, the extractor is able to extract $r=\frac{z_{r_j}-z'_{r_j}}{\beta_1-\beta_2}$. By setting $z_r=r$, the simulator can compute $g^{z_{x_j}} \cdot h^r = Cm(x'_j,0) \cdot \frac{Cm(x, r)}{g^{x_j}}$. Finally, for both $j\in\{1,...,n\}\backslash\{i\}$ and $j=i$, the simulator can get $x=x_i\in \mathcal X$. 
    \item ({\tt zkpNN}): The extractor sends two random challenges $\beta_1$ and $\beta_2$ before and after rewinding, and the prover replies with $z_r = r' + \beta_1 \cdot (\sum_{i=1}^m r_i \cdot 2^{i-1} - r)$ and $z'_r = r' + \beta_2 \cdot (\sum_{i=1}^m r_i \cdot 2^{i-1} - r)$, accordingly. Next, the extractor is able to extract $\sum_{i=1}^m r_i \cdot 2^{i-1} - r$ by computing $\frac{z_r-z_r'}{\beta_1-\beta_2}$. By setting $z_r=\sum_{i=1}^m r_i \cdot 2^{i-1} - r$, the simulator is able to get $h^{z_r}=Cm(x,r)^{-1} \cdot \prod_{i=1}^m Cm(b_i, r_i)^{2^{i-1}}$, and hence $x=\sum_{i=1}^{m}b_i \cdot 2^{i-1} \geqslant 0$.
\end{itemize}

\subsection{Honest Verifier Zero-knowledge Proof} \

To prove zero-knowledge, we need to construct a simulator, which interacts with the verifier and can eventually convince the verifier by simulating the transcript of the proof like a prover. However, the simulator does not know the knowledge throughout the interaction.  

\begin{itemize}
    \item ({\tt zkpCm}): The simulator first sends some initial message to the verifier so that the random challenge $\beta$ could be obtained. Next, the simulator rewinds the verifier, randomly generates $(x', r') \in {\mathbb Z}^2_p$, and send $Cm(x,r)^{-\beta}\cdot Cm(x',r')$ to the verifier as the initial message. Once the verifier challenges on $\beta$ again, the simulator replies with $z_x=x'$ and $z_r=r'$.  
    \item ({\tt zkpMbs}): Similarly, the simulator first randomly sends some initial message to the prover to get the challenge $\beta$, and then rewinds the verifier. Next, the simulator randomly picks $x'=x'_i \in {\mathcal X}$, randomly generates $(x'_j, r'_j) \in {\mathbb Z}_p^2$ and $\beta_j$, and computes $\beta_i'$ and $z_{r_j}$, according to the real prover. In addition, the simulator sets $Cm(x'_j,r'_j) = g^{z_{x_j}} \cdot h^{z_{r_j}} \cdot \Big( \frac{Cm(x, r)}{g^{x_j}} \Big)^{-\beta_j}$ for all $j$. Finally, the simulator sends  $(Cm(x'_j, r'_j), z_{x_j})_{j=1}^n$ to the verifier, and once it challenges on $\beta$ again, the simulator replies with $(\beta_j, z_{r_j})_{j=1}^n$.
    \item ({\tt zkpNN}): The simulator randomly send messages in step 1 to get the challenge $\beta$, and then rewinds the verifier. Next, the simulator randomly generates $m, b_i, r_i, r',r''$ and set $z_r = r' + \beta \cdot (\sum_{i=1}^m r_i \cdot 2^{i-1} - r'')$ and $Cm(0,r')= h^{z_r} \cdot Cm(x, r)^{\beta} \cdot \prod_{i=1}^m Cm(b_i, r_i)^{-\beta\cdot 2^{i-1}}$. Finally, the simulator completes step 1 as the real prover, using the values generated, and when the verifier challenges it on $\beta$ again, the simulator replies with $z_r$.
\end{itemize}

\end{document}